\documentclass[12pt]{article}
\usepackage{latexsym,epsf}
\usepackage{epsfig,amsfonts,cite}

\makeatletter
\@addtoreset{equation}{section}

\makeatother     

\begin{document}

\title{Testing the efficiency of different improvement programs}

\author{
  \\
  {\small Sergio Caracciolo and Andrea Montanari}              \\[-0.2cm]
  {\small\it Scuola Normale Superiore and INFN -- Sezione di Pisa}  \\[-0.2cm]
 {\small\it I-56100 Pisa, ITALIA}          \\[-0.2cm]
  {\small Internet: {\tt sergio.caracciolo@sns.it, montanar@cibs.sns.it}}   
  \\[-0.2cm]
  \\[-0.1cm]  \and
  {\small Andrea Pelissetto}              \\[-0.2cm]
  {\small\it Dipartimento di Fisica and INFN -- Sezione di Roma I}    \\[-0.2cm]
  {\small\it Universit\`a degli Studi di Roma ``La Sapienza"}        \\[-0.2cm]
  {\small\it I-00185 Roma, ITALIA}          \\[-0.2cm]
  {\small Internet: {\tt pelisset@ibmth.df.unipi.it}}   \\[-0.2cm]
  {\protect\makebox[5in]{\quad}}  
  \\
}
\vspace{0.5cm}


\def\sg{\mbox{\boldmath $\sigma$}}
\def\s#1{\mbox{\boldmath $\sigma$}_{#1}}
\def\g{\gamma}
\def\p{\hat{p}}
\def\b{\beta}
\def\sgp#1{\mbox{\boldmath$\sigma$}_{\| #1}}
\def\sgn#1{\mbox{\boldmath$\sigma$}_{\bot #1}}
\def\nr{\mbox{\boldmath$n$}}
\def\e{\epsilon}
\def\is{\sigma}
\def\o{\omega_L}
\def\oi{\omega_{\infty}}
\def\a#1{\hat{\alpha}_{#1}}
\def\ah{\hat{\alpha}}
\def\m{\overline{p}}
\def\ph{\mbox{\boldmath $\phi$}}

\newcommand\D{\displaystyle}


\newcommand{\be}{\begin{equation}}
\newcommand{\ee}{\end{equation}}
\newcommand{\<}{\langle}
\renewcommand{\>}{\rangle}

\def\spose#1{\hbox to 0pt{#1\hss}}
\def\ltapprox{\mathrel{\spose{\lower 3pt\hbox{$\mathchar"218$}}
 \raise 2.0pt\hbox{$\mathchar"13C$}}}
\def\gtapprox{\mathrel{\spose{\lower 3pt\hbox{$\mathchar"218$}}
 \raise 2.0pt\hbox{$\mathchar"13E$}}}

\def\bsigma{\mbox{\protect\boldmath $\sigma$}}
\def\bpi{\mbox{\protect\boldmath $\pi$}}
\def\btau{\mbox{\protect\boldmath $\tau$}}
\def\br{\mbox{\protect\boldmath $r$}}
\def\hatp{\hat p}
\def\hatl{\hat l}

\def\msbar{ {\overline{\hbox{\scriptsize MS}}} }
\def\normalmsbar{ {\overline{\hbox{\normalsize MS}}} }

\newcommand{\R}{\hbox{{\rm I}\kern-.2em\hbox{\rm R}}}

\newcommand{\reff}[1]{(\ref{#1})}
\def\smfrac#1#2{{\textstyle\frac{#1}{#2}}}


\maketitle
\thispagestyle{empty}   

\vspace{0.2cm}

\begin{abstract}
We study the finite-size behaviour of a tree-level on-shell
improved action for the $N$-vector model. We present numerical
results for $N=3$ and analytic results in the large-$N$ limit for the 
mass gap. We also report a perturbative computation at one loop
of the mass gap for states of spatial momentum $p$.
We present a detailed comparison of the behaviour of this action with that of
other formulations, including the perfect action, and a critical discussion
of the different approaches to the problem of action improvement.
\end{abstract}

\clearpage

%
%
\section{Introduction}

Lattice simulations are at present the most effective method to investigate 
non-perturbative properties of field theories like QCD. By
necessity Monte Carlo simulations are performed
on finite lattices and at finite values of the correlation length.
It is therefore of utmost importance to understand the systematic effects
due to scaling corrections.

In the last years a lot
of work has been devoted to invent lattice models that have  small
scaling corrections  so that continuum results can be obtained on small
lattices and thus with a limited use of computer time.

This program was started by Symanzik \cite{Symanzik} who put up
a systematic method to improve asymptotically free theories using 
perturbation theory. Soon after, L\"uscher and Weisz noticed that simpler 
actions could be used if one was only interested in on-shell quantities
\cite{LuschShell} .
A different approach is the {\em perfect-action} program  
in which improved actions are determined as fixed points of
renormalization-group transformations 
\cite{HasenPer,Hasenfratz_94_98,DeGrand-etal,Bietenholz-etal,%
Orginos-etal,Farchioni-etal,Spiegel}. By definition classically
perfect actions do not show any lattice effect at tree level.
In the standard language they are Symanzik tree-level on-shell improved 
to all orders of $a$ \cite{Hasenfratz-Niedermayer_97}. 

Recently the improvement program has been implemented non-perturbatively
for the fermionic action
\cite{Luscher_LAT97,Edwards_LAT97,Gockeler_LAT97,Dawson_LAT97}.
This represents an important step forward. Indeed actions that are 
improved to a finite number of loops have scaling corrections of 
order $O(a \log^p a)$, or, in the statistical mechanics language,
of order $O(\xi^{-1} \log^p \xi)$, where $\xi$ is the 
correlation length. On the other hand non-perturbatively improved actions 
should have corrections of order $O(\xi^{-2} \log^q \xi)$. It must 
be noted however that improvement comes with a price. Improved 
actions are more complicated than unimproved ones. Thus to understand 
the practical relevance of any improvement one should also consider 
the additional cost in the simulations.

In this paper we will study the two-dimensional $N$-vector model.
This theory provides the simplest example
for the realization of a nonabelian global symmetry.
Its two-dimensional version has been
extensively studied because it shares with four-dimensional gauge
theories the property of being asymptotically free in the
weak-coupling perturbative
expansion~\cite{Polyakov_75,Brezin_76,Bardeen_76}.
This picture predicts a nonperturbative generation of a mass gap that
controls the exponential decay of the correlation
functions at large distance.

Besides perturbation theory, the two-dimensional
$N$-vector model can be studied using different techniques.
It can be solved in the $N=\infty$ limit~\cite{Stanley,DiVecchia}
and $1/N$ corrections can be systematically
calculated~\cite{Muller,Flyvbjerg,Campostrini_90ab}.
An exact $S$-matrix can be computed~\cite{Zamolodchikov_79,
Polyakov-Wiegmann_83}
and, using the thermodynamic Bethe ansatz, the exact mass gap
of the theory in the limit $\beta\to\infty$ has been obtained
\cite{Hasenfratz-Niedermayer_1,Hasenfratz-Niedermayer_2}.
The model has also been the object of extensive numerical work
\cite{Wolff_O4_O8,MGMC_O4,o3_letter,CEMPS,Alles-Symanzik,Alles-e-turchi}
mainly devoted to checking the
correctness of the perturbative predictions
\cite{Falcioni-Treves,CP-3loop,CP-4loop,Shin}.

We will consider the action that has been proposed 
in Ref. \cite{Caracciolo-Pelissetto_97}. It is on-shell 
tree-level improved and satisfies reflection positivity 
so that a positive transfer matrix can be defined. Moreover 
we will show here that this action can be efficiently 
simulated: indeed one can use a Wolff algorithm 
\cite{Wolff_89_90,Edwards-Sokal_89,Hasenbusch_90,Caracciolo-etal_93}
with unfrustrated embedded Ising model. Therefore no critical slowing
down is expected. This is at variance with the 
standard Symanzik tree-level action. Indeed, also in this case 
one can define a cluster algorithm \cite{Buonanno-Cella_95}.
However the embedded Ising model is 
frustrated and therefore critical slowing down is still present.

The purpose of this work is twofold. First of all we want to understand
quantitatively the effect of tree-level improvement. Indeed it is not
obvious {\em a priori} that the idea is effective since the corrections
to scaling are simply reduced a logarithm of $a$. As we shall see from 
our numerical results for $N=3$ and our analytic expressions in the 
large-$N$ limit, tree-level improvement effectively works: 
indeed one finds ``naturally" --- i.e. without any additional 
tuning of the parameters --- a reduction of the scaling corrections
by approximately a factor of two. The second purpose of this work 
was to understand the results of Ref. \cite{HasenPer} for the perfect 
action which shows, for a particular value of the parameter $\kappa$
that parametrizes the renormalization-group transformations, a dramatic 
improvement with respect to the standard action. The on-shell action
that is considered in this paper differs from the action of 
Ref. \cite{HasenPer} by terms of order $a^4$ and it shares the property 
of being extremely local. We thus wanted to understand if the results of 
Ref. \cite{HasenPer} depended only on the improvement and on the locality
of the action. The answer is clearly negative, since the on-shell action we 
study has much larger corrections. At this point the question that 
arises naturally is whether the exceptionally good behaviour is related 
to the fact
that the action is a classical fixed point of a renormalization-group
transformation. Extending the perturbative calculation of Ref.
\cite{Hasenfratz-Niedermayer_97} we will see that this interpretation 
is unlikely. Indeed with a different choice of the parameter $\kappa$
one can define perfect actions that are local but that are not 
expected to have such a good behaviour.

In this paper we will investigate the corrections to 
finite-size scaling (FSS) for various models that are
introduced in Sect. \ref{sec2}. 
A discussion of the FSS corrections in the large-$N$
limit is presented in Sect. \ref{sec3}. We will show here that 
tree-level improvement reduces the corrections to scaling 
from $\log L/L^2$ to $1/L^2$. Sect. \ref{sec4} presents our algorithm
and Monte Carlo results. Finally in Sect. \ref{sec5} we present our
conclusions and compare our results with those obtained with other 
types of action. App. \ref{AppA} presents the details of the 
analytic calculation of the FSS functions for the on-shell action in the 
large-$N$ limit, App. \ref{AppB} the analytic computation 
at one loop of the mass gap for non-vanishing spatial momentum
while in App. \ref{AppC} we give some details of our calculations with
the perfect action.

Preliminary results of this work have been presented at the Lattice 97
conference \cite{Caracciolo-etal_98}.

\section{The models} \label{sec2}

In this work we will study in detail the FSS properties of the 
tree-level on-shell improved action proposed 
in Ref. \cite{Caracciolo-Pelissetto_97}:
\begin{eqnarray}
S^{on shell}(\sg)&=&
    \sum_x\left[ \frac{2}{3}\sum_{\mu}(\s{x}\cdot\s{x+\mu})
     +\frac{1}{6}\sum_{d}(\s{x}\cdot\s{x+d})\right.
\nonumber \\
&& \left. 
-\frac{1}{24}\sum_{s_1=\pm 1} \sum_{s_2=\pm 1}
   \left(\s{x}\cdot\s{x+s_1 \hat{1}} + \s{x}\cdot\s{x+s_2 \hat{2}} -2
         \right)^2  \right],
\label{ons}
\end{eqnarray}
where $d$ runs over the diagonal vectors $(1,\pm 1)$, while in the last 
term $\hat{1}$ and $\hat{2}$ are the unit vectors along the $x$- and 
$y$-axis respectively. It was shown in Ref. \cite{Caracciolo-Pelissetto_97}
that the action \reff{ons} is reflection-positive and tree-level
on-shell improved. In the formal continuum limit we have
\be
S^{on shell}(\sg)\approx 
   \int d^2 x \left[ {1\over2} (\bsigma\cdot \Box\bsigma) +
  {\alpha_2 a^2\over2} (\bsigma\cdot \Box^2 \bsigma -
      (\bsigma\cdot \Box \bsigma)^2 ) + O(a^4) \right]
\label{Honshellimp}
\ee
where $\Box = \sum_\mu \partial_\mu^2$ and $\alpha_2 = -{1\over12}$.

We will study the FSS corrections for this action and we will compare
these results with those obtained for other 
actions that have been extensively studied in the literature:
\begin{itemize}
\item the {\it standard action}
\be
S^{std}(\sg) =\, \sum_{x,\mu}\s{x}\cdot\s{x+\mu}
\label{std}
\ee
whose FSS behaviour has been extensively studied in Refs. 
\cite{LuschNum,o3_letter,Shin_97};
\item the {\it Symanzik action} \cite{Symanzik}
\be
S^{sym}(\sg) =\, \sum_{x,\mu}\left(\frac{4}{3}\s{x}\cdot\s{x+\mu}-
\frac{1}{12}\s{x}\cdot\s{x+2\mu}\right)
\label{sym}
\ee
which is designed to cancel $O(a^2)$ lattice artifacts in tree-level
Green functions \cite{Symanzik};
\item the {\it diagonal action} \cite{Niedermayer_LAT96}
\be
S^{diag}(\sg)=\sum_{x,d}\left(\frac{2}{3}\s{x}\cdot\s{x+\mu}+
\frac{1}{6}\s{x}\cdot\s{x+d}\right)
\label{dia}
\ee
whose one-particle spectrum has no $O(a^2)$ artifacts at tree level
but that is not on-shell improved. For instance the four-point 
function shows $O(a^2)$ corrections even on-shell.
\item the classically perfect action \cite{HasenPer,Hasenfratz_94_98}
that is defined as the fixed point of a class of 
renormalization-group transformations for $\beta=\infty$. 
The two-spin part is given by
\be
S^{per\!f,2spin}(\sg) = 
   \sum_{x,y} w(x-y) \s{x}\cdot\s{y},
\ee
where $w(x)$ is the so-called {\em perfect laplacian} \cite{Bell-Wilson_74}.
Its Fourier transform is given by
\be
{1\over w(q)}\, =\, 
{1\over 3\kappa}+\, 
   \sum_{l_1,l_2=-\infty}^\infty 
   {1\over (q_1 + 2 \pi l_1)^2 + (q_2 + 2 \pi l_2)^2}\,
   {\widehat{q}_1^2 \widehat{q}_2^2\over 
    (q_1 + 2 \pi l_1)^2 (q_2 + 2 \pi l_2)^2},
\label{perfect-laplacian} 
\ee
where $\kappa$ is a parameter that characterizes the renormalization-group
transformation. The four-spin coupling --- and also higher order couplings ---
cannot be computed in closed form. In App. \ref{AppC} we give 
some details on our determination of the four-spin term 
for various values of $\kappa$. The formal continuum limit of the 
perfect action is given by Eq. \reff{Honshellimp} with 
$\alpha_2 = (\kappa-4)/(12 \kappa)$. Notice that 
the perfect action with $\kappa = 2$ and the on-shell action \reff{ons} 
are, at tree level, equivalent also off-shell up to terms of order $a^4$.
\end{itemize}
%
%
\section{FSS functions in the large-$N$ limit} \label{sec3}

In this section we will discuss the relation between improvement 
and finite-size scaling in the large-$N$ limit extending the discussion
of Ref. \cite{Caracciolo-Pelissetto_98}. We will 
consider an $L\times T$ square lattice and we will generalize
the action (\ref{ons}) considering
\begin{equation}
S[\sg]=N\sum_{x,y}J(x-y)\s{x}\cdot
\s{y} + \frac{\alpha_3 N}{2}\sum_{x,y,z}K(x-y)K(x-z)(\s{x}\cdot
\s{y})(\s{x}\cdot \s{z}),
\label{azione-generale}
\ee
where 
\begin{eqnarray}
\hat{K}(p)& \equiv &\sum_x K(x)e^{-ipx}= p^2+O(p^4), \\ 
\hat{J}(p)& \equiv &\sum_x J(x)e^{-ipx}
\equiv\hat{J}(0)-\frac{1}{2}w(p),\\
w(p)&=&\p^2+\alpha_1\sum_{\mu}\p_{\mu}^4+\alpha_2 (\p^2)^2+O(p^6).
\end{eqnarray}
Here $\p^2\equiv\sum_{\mu}\p_{\mu}^2\equiv
\sum_{\mu}(2\sin(p_{\mu}/2))^2$. 

We will assume, as usual, that all the couplings respect the symmetry of the 
lattice and that $w(p)$ vanishes only for $p=0$ in the Brillouin zone. 
Moreover we require the action to be ferromagnetic, that is to have 
a unique maximum corresponding to the ordered configuration. 
The general class of actions \reff{azione-generale} was studied in
Ref. \cite{Caracciolo-Pelissetto_97} where it was shown
that $S[\sg]$ is tree-level on-shell improved if $\alpha_1 = 1/12$
and $\alpha_2 = \alpha_3$.

In the large-$N$ limit the model can be solved using 
a standard Lagrange-multiplier technique. One introduces two parameters
$m_{L,T}$ and $\omega_{L,T}$ related to $\beta$ by the 
gap equations\footnote{We should remark that for some choices of $K(x)$
and $J(x)$ these equations may not be correct for all values of $\beta$
(see for instance the exact solution of the mixed 
$O(N)-RP^{N-1}$ models of Ref. \cite{Magnoli-Ravanini}). However under our 
hypotheses they should always describe the theory for sufficiently large 
values of $\beta$. This is sufficient for our analysis since 
we are only interested in the critical limit $\beta\to\infty$.
}:
\begin{eqnarray}
\b (1+\omega_{L,T}) &=& {1\over LT} \sum_{p}
\frac{1}{\hat{w}(p;\omega_{L,T}) +m^2_{L,T}} ,
\label{gap1} \\
-\frac{\b}{2\alpha_3}(1+\omega_{L,T})\omega_{L,T} &= &
{1\over LT} \sum_{p} \frac{\hat{K}(p)}{\hat{w}(p;\omega_{L,T})+m^2_{L,T}} ,
\label{gap2} 
\end{eqnarray}
where
\begin{equation}
\hat{w}(p;\omega ) \equiv \frac{w(p)+\omega \hat{K}(p)}{1 + \omega} .
\label{dp}
\end{equation}
The two-point function is simply given by:
\begin{equation}
\langle\s{x} \cdot \s{y}\rangle  =  \frac{1}{\b(1+\omega_{L,T})}
 {1\over LT} \sum_p
\frac{e^{ip(x-y)}}{\hat{w}(p;\omega_{L,T}) +m^2_{L,T}}.
\label{corr}
\end{equation}
In Appendix \ref{AppA} 
we present an analytic computation of the FSS curves and 
of their leading correction. For the mass gap $\mu(L,\beta)$ defined on a strip
$L\times \infty$, in the limit $\beta\to\infty$, $L\to\infty$,
with $\mu(L,\beta) L\equiv x$ fixed, we find 
\begin{equation}
\left( {\mu(\infty,\b)\over \mu(L,\b)} \right)^2 \ =\, 
   f_\mu (x) \left(1 + {\Delta_\mu(x;L)\over L^2} + 
    O(L^{-4} \log^2 L)\right).
\label{FSSexpansion}
\end{equation}
The function $f_\mu(x)$ is the FSS function and it was already
computed by L\"uscher~\cite{Luscher_81}:
\be
f_\mu(x)= \exp\left[-4\sum_{n=1}^{\infty}K_0(nx)\right] = 
\frac{16\pi ^2 e^{-2\gamma_E}}{x^2}\exp\left[-\frac{2\pi}{x}\right](1+O(x)) .
\label{FSSfunction}
\ee
The function $\Delta_\mu(x;L)$ is the leading correction to FSS. For 
$L\to\infty$ keeping $x$ fixed, 
it behaves as $\log L$ and it has a regular expansion in 
powers of $1/\log L$ of the form
\begin{equation}
\Delta_\mu(x;L) = \sum_{q=-1}^\infty \delta_{\mu,q}(x) (\log L)^{-q}.
\label{expansionDelta}
\end{equation}
The leading term is given by
\be
\delta_{\mu,-1} (x) =\
    \frac{x^2}{4}(1-f_\mu(x))(1-12\alpha_1-16\alpha_2+16\alpha_3).
\ee
It vanishes for tree-level improved actions. Thus improvement 
reduces the corrections to scaling from $\log L/L^2$ to $1/L^2$.
Generically, the next coefficient $\delta_{\mu,0}(x)$ does not vanish 
for tree-level improved actions. However one can choose $K(x)$ 
and $J(x)$ in order to have $\delta_{\mu,0}(x)=0$ (see Appendix \ref{AppA}). 
In other words, with an appropriate choice of these two functions it is 
possible to improve the action to one-loop order. One can 
verify that it is not possible to obtain $\delta_{\mu,1}(x) = 0$ for 
$\alpha_3\not=0$. In this case, actions of the 
form \reff{azione-generale} cannot be two-loop on-shell improved.
This is not unexpected since, six-spin
couplings are needed for two-loop improvement. 
If $\alpha_3=0$ the situation is simpler.
Indeed in this case, once the action is one-loop improved, it is automatically
improved to all orders of perturbation theory \cite{Caracciolo-Pelissetto_97}.
This is however an accident of the large-$N$ limit and it is not true for 
finite values of $N$.
\begin{table}
\begin{center}
\begin{tabular}{|l|rrr|}
\hline
 $L$ &  $S^{std}$  &  $S^{sym}$  &   $S^{onshell}$   \\
\hline
4  & 0.79362345    & 0.78565020  & 0.78324283  \\
6  & 0.78640181    & 0.78100623  & 0.78024309  \\
8  & 0.78290367    & 0.77913556  & 0.77884796  \\
10 & 0.78099463    & 0.77822405  & 0.77808048  \\
16 & 0.77857473    & 0.77720115  & 0.77717924  \\
20 & 0.77792482    & 0.77696043  & 0.77697663  \\
32 & 0.77714433    & 0.77669433  & 0.77669910  \\
\hline
$\infty$ 
   & 0.77652331    & 0.77652331  & 0.77652331  \\       
\hline
\end{tabular}
\end{center}
\caption{$\mu(\infty,\beta)/\mu(L,\beta)$ for $x=2$ for different lattice
actions.  }
\label{tableFSS1}
\end{table}

Since tree-level improvement reduces FSS corrections only by a
logarithm of $L$, one could be skeptical on the effectiveness of the 
idea. We have therefore compared the FSS scaling corrections for various
actions. In Table \ref{tableFSS1} 
we report $\mu(\infty,\beta)/\mu(L,\beta)$ for various 
values of $L$ for $x=2$ for the on-shell action 
(\ref{ons}), as well as for the actions \reff{std} and \reff{sym}.
It is clear that tree-level improved 
actions show much smaller corrections. For the purpose of future 
comparison we consider also the quantity 
\be
  R(L,x) \equiv 2 L \mu(2 L,\beta) .
\label{RLxdef}
\ee
It is easy to convince oneself that in the FSS limit $R(L,x)$ assumes 
a finite value $R(\infty,x)$ with corrections of order $\log L/L^2$ . 
In Fig. \ref{RNinf} we show $R(L,x)$ for $x = 1.0595$ for various actions. 
Also here it is evident that improved actions show smaller corrections to
scaling.

Let us now discuss the validity of the expansion \reff{FSSexpansion}. 
First of all the expansion is not uniform in $x$: the error increases as 
$x\to \infty$. This fact can be checked analytically from the 
exact expressions of App. \ref{AppA}. Its origin is easy to understand. 
In order to have FSS one should work at values of $\beta$ such that 
the correlation length is much larger than one lattice spacing,
i.e. $\mu(L) \ll 1$. This means that the expansion \reff{FSSexpansion}
makes sense only for $x/L \ll 1$.

We want finally to discuss the connection between the FSS limit 
for $x\to 0$ and perturbation theory (PT). The PT limit 
corresponds to $\beta\to\infty$ with $L$ fixed.  
For instance for $\mu(L)$ one obtains
\be
x \equiv \mu(L) L = {1\over \beta} \sum_{n=0}^\infty 
   {a_n (L)\over \beta^n},
\ee
where $a_n(L)$ has an expansion of the form
\be
a_n(L) = \sum_{k=0}^n a_{nk}^{(0)} \log^k L + 
      {1\over L^2} \sum_{k=0}^n a_{nk}^{(1)} \log^k L + O(L^{-4}).
\label{expansion-an}
\ee
This expansion can be inverted to give
\be
\beta = {1\over x} \sum_{n=0}^\infty {b_n(L)x^n},
\ee
where the coefficient $b_n(L)$ have an expansion of the form 
\reff{expansion-an}. To obtain the FSS curve one can then
use the asymptotic freedom prediction 
\be
\mu(\infty) = A \left({2 \pi N \beta \over N-2}\right)^{1\over N-2}
    \exp\left(-{2 \pi N \beta\over N-2}\right)
    \left(1 + \sum_{n=1}^\infty {c_n\over \beta^n}\right),
\ee
and substitute the expansion of $\beta$ in terms of $x$. In this way
one computes $\mu(\infty)/\mu(L)$. The leading term for $L\to\infty$ 
is $L$-independent and correctly reproduces the expansion of the 
exact result \reff{FSSfunction}. One can also compute the first 
correction in powers of $1/L^2$. In this way one obtains 
the expansion of $\Delta_\mu(x;L)$ for $L$ fixed in the limit
$x\to0$:
\be
\Delta_\mu(x;L) = \sum_n P_n(\log L) x^n,
\ee
where $P_n(x)$ is a degree-$n$ polynomial.
Clearly this expansion is incorrect for $L\to\infty$, although,
for finite small $x$, it gives a reasonable approximation as long as 
$L\ll L_{max}(x) \approx e^{\pi/x}$. In the opposite limit 
$L\gg L_{max}(x)$, the 
correct expansion is given by Eq. \reff{expansionDelta}. 
This result reflects the fact that the PT limit $\beta\to\infty$ 
at fixed $L$ (equivalent to $x\to0$ at fixed $L$), followed 
by $L\to \infty$ does not commute with the FSS limit followed 
by $x\to 0$, except for the leading term \cite{Caracciolo-Pelissetto_98}.

Numerically PT calculations of the corrections to FSS can still be 
useful, at least if one is able to compute enough terms. 
We have considered for instance $R(L,x)$ that admits an expansion of 
the form 
\be
R(L,x) = R_0(x) + {1\over L^2} R_1(x;\log L) + \ldots 
\ee
where $R_0(x) \equiv R(\infty,x)$ is the FSS function. For the 
on-shell action \reff{ons} we have up to three-loops
\be
R_1(x;\log L) = 
-0.0251916 x^3 + (0.0321612 + 0.008018747\log L) x^4 + O(x^5).
\ee
Because the action is improved there is no $x^2$ (one-loop) term.
Consider now $\Delta R(L,x) \equiv R(L,x) - R_0(x)$. For $x=1.0595$
and $L=8,64$ we obtain the estimates
\begin{eqnarray*}
 \Delta R(8,1.0595) &=& 
    0 \; (\hbox{\rm 1 loop}); \quad -0.4681\cdot 10^{-3} \; (\hbox{\rm 2 loops}); 
               \quad 0.4934\cdot 10^{-3} \; (\hbox{\rm 3 loops}), \\
 \Delta R(64,1.0595) &=& 
    0 \; (\hbox{\rm 1 loop}); \quad  -0.7314\cdot 10^{-5} \; (\hbox{\rm 2
    loops});
              \quad 0.1284\cdot 10^{-4}\;  (\hbox{\rm 3 loops}).
\end{eqnarray*}
This should be compared with the exact  results $0.751\cdot 10^{-3}$
for $L=8$ and 
$0.147\cdot 10^{-4}$ for $L=64$. The three-loop result is reasonably close to
the exact value while the two-loop approximation gives a grossly inexact guess 
(it has even the wrong sign) of the exact result.

%
%
\section{Monte Carlo results} \label{sec4}

As we have seen in the previous section tree-level improvement 
changes the correction to FSS by a $\log L$. A priori this appears 
to be a very small improvement. However at $N=\infty$ 
the actions \reff{ons} and \reff{sym} show a much better
behaviour with respect to the standard action \reff{std}. We decided 
to check if this is true also for $N=3$, comparing our 
data for the mass gap with the simulation results of Ref.
\cite{LuschNum} for the standard action
and of Ref. \cite{HasenPer} for the perfect action. 

We have used  in our simulation the Wolff algorithm 
\cite{Wolff_89_90,Edwards-Sokal_89,Hasenbusch_90,Caracciolo-etal_93}
with standard Swendsen-Wang updatings
\cite{Swendsen-Wang,Edwards-Sokal_88}.
The idea is the following: given a spin configuration
$\{\bsigma_x\}$, one chooses randomly a unit vector $\br$
and defines Ising variables $\epsilon_x$ rewriting
\be
\bsigma_{x} = \epsilon_x \sgp{x} + \sgn{x},
\label{embedding}
\ee
where $\sgp{x}$ and $\sgn{x}$ are the components of $\bsigma_x$ 
respectively parallel and perpendicular to $\br$. 
Eq.  \reff{embedding} defines an effective action 
$S^{\hbox{\it eff}}(\{\epsilon\})$ for the embedded Ising variables. A simplifying
feature of our action \reff{ons} is that the effective 
action $S^{\hbox{\it eff}}(\{\epsilon\})$ contains only two-spin Ising couplings
and therefore it can be updated using a standard cluster algorithm.
We can indeed rewrite the action \reff{ons} as
\setlength{\unitlength}{1cm}
\begin{center}
\begin{picture}(15.0,3.0)
\put(-0.3,2.){$S^{on shell}=\ \sum_x \Bigg\{ 
\frac{4}{3}\Bigg($}
\put(3.2,1.6){\line(0,1){1}}\put(3.5,2.){+}\put(4.2,2.){\line(1,0){1}}
\put(3.2,1.6){\circle*{0.1}}\put(3.2,2.6){\circle*{0.1}}
\put(4.2,2.0){\circle*{0.1}}\put(5.2,2.0){\circle*{0.1}}
\put(5.3,2.){$\Bigg)+\ \frac{1}{6}\Bigg($}
\put(6.85,2.4){\line(1,-1){0.7}}\put(7.8,2.){+}
\put(6.85,2.4){\circle*{0.1}}\put(7.55,1.7){\circle*{0.1}}
\put(8.3,1.7){\line(1,1){0.7}}\put(9.0,2.){$\Bigg) +$}
\put(8.3,1.7){\circle*{0.1}}\put(9.0,2.4){\circle*{0.1}}
\put(1.7,0.6){$-\frac{1}{24}\Bigg[\Bigg($}
\put(2.9,0.6){\line(1,0){1}}\put(2.9,0.65){\line(1,0){1}}
\put(2.9,0.625){\circle*{0.1}}\put(3.9,0.625){\circle*{0.1}}
\put(4.1,0.6){+}
\put(4.6,0.2){\line(0,1){1}}\put(4.65,0.2){\line(0,1){1}}
\put(4.625,0.2){\circle*{0.1}}\put(4.625,1.2){\circle*{0.1}}
\put(4.8,0.6){+ 2}
\put(5.8,0.2){\line(0,1){1}}\put(5.8,0.2){\line(1,0){1}}
\put(5.8,0.2){\circle*{0.1}}\put(5.8,1.2){\circle*{0.1}}
\put(6.8,0.2){\circle*{0.1}}
\put(7,0.6){$\Bigg)+$ .... $\Bigg]\Bigg\}$}
\end{picture}
\end{center}
\noindent 
The Wolff embedding acts on these couplings as follows: 
\begin{eqnarray}
\begin{picture}(3,.5)
\put(0.,.1){\line(1,0){2.}}
\put(0.,.1){\circle*{0.2}}\put(2.0,.1){\circle*{0.2}}
\put(0.,.5){x}\put(2.0,.5){y}
\end{picture}
&=& \sg_x\cdot\sg_y  \to 
\e_x\e_y\sgp{x}\cdot\sgp{y}\\
\nonumber\\
\begin{picture}(3,.5)
\put(0.,.05){\line(1,0){2.}}
\put(0.,.15){\line(1,0){2.}}
\put(0.,.1){\circle*{0.2}}\put(2.0,.1){\circle*{0.2}}
\put(0.,.5){x}\put(2.0,.5){y}
\end{picture}
& = & (\sg_x\cdot\sg_y)^2\to
2\e_x\e_y\sgp{x}\cdot\sgp{y}\sgn{x}\cdot\sgn{y}\\
\nonumber\\
\begin{picture}(3,1.5)
\put(0.,1.1){\line(1,0){2.}}
\put(0.,1.1){\circle*{0.2}}\put(2.0,1.1){\circle*{0.2}}
\put(0.,1.5){y}\put(2.0,1.5){z}
\put(0.,-.9){\line(0,1){2.}}
\put(0.,-0.9){\circle*{0.2}}
\put(0.2,-0.5){x}
\end{picture}
& = & (\sg_y\cdot\sg_x)(\sg_y\cdot\sg_z)  \to 
\e_x\e_z\sgp{x}\cdot\sgp{y}\sgp{y}\cdot\sgp{z}\\
&& \hphantom{ (\sg_y\cdot\sg_x)(\sg_y\cdot\sg_z)  \to }
+\e_y\e_z\sgp{y}\cdot\sgp{z}\sgn{x}\cdot\sgn{y}\nonumber\\
&& \hphantom{ (\sg_y\cdot\sg_x)(\sg_y\cdot\sg_z)  \to }
+\e_y\e_x\sgp{y}\cdot\sgp{x}\sgn{z}\cdot\sgn{y}\nonumber
\end{eqnarray}
The embedded action becomes
\be
S^{\hbox{\it eff}}(\{\epsilon\}) = 
   \sum_x \left[ \sum_\mu J^{(1)}_{x\mu} \epsilon_x \epsilon_{x+\mu} +
                 \sum_d J^{(2)}_{xd} \epsilon_x \epsilon_{x+d}\right],
\label{effHam}
\ee
where $d$ are the diagonal vectors $(1,\pm1)$. The couplings are defined 
by 
\begin{eqnarray}
J^{(1)}_{x\mu} &=& (\sgp{x}\cdot\sgp{x+\mu}) 
    \left[{4\over3} - {1\over 3} \sgn{x}\cdot \sgn{x+\mu} - 
      {1\over12} \sgn{x} \cdot \sgn{x+\nu} -
      {1\over12} \sgn{x} \cdot \sgn{x-\nu} -  \right.
\nonumber \\
  && \qquad \left. {1\over12} \sgn{x} \cdot \sgn{x+\nu+\mu} -
                   {1\over12} \sgn{x} \cdot \sgn{x-\nu+\mu}
      \right],
\\
J^{(2)}_{xd} &=& = {1\over6} (\sgp{x} \cdot \sgp{x+d}) 
   \left[1 - {1\over2} \sgp{x+\hat{1}}^2 - {1\over2} 
     \sgp{x+d-\hat{1}}^2\right],
\end{eqnarray}
where, in the first definition, $\nu$ is a unit vector orthogonal to $\mu$.
It is easy to see that the embedded Ising model is not frustrated. 
Indeed if we redefine the Ising variables as 
\be 
{\epsilon'}_x = \epsilon_x\, {\rm sign}\,(\bsigma\cdot \br)
\ee
we obtain a new effective action of the type \reff{effHam}
with $J^{(1)}_{x\mu} \ge 0$, $J^{(2)}_{xd} \ge 0$. 
Therefore we expect the algorithm  not to show any critical slowing down
\cite{Caracciolo-etal_93}.

The purpose of our simulation was the determination of the FSS function
$R(L,x)$ defined in Eq. \reff{RLxdef} for 
$x\equiv \mu(L,\beta) L = 1.0595$ for a strip 
$L\times \infty$ with periodic boundary conditions (pbc) in the 
spatial direction. This is indeed the value of $x$ 
for which results are available for the standard 
\cite{LuschNum} and for the perfect action \cite{HasenPer}.
Given $L$, in order to compute $R(L,x)$, we have first determined 
$\tilde{\beta}$ such that $\mu(\tilde{\beta},L) = 1.0595$. 
This has been obtained performing a Monte Carlo simulation 
at a nearby value $\beta_{run}$ of $\beta$, and appropriately reweighting the 
correlation functions. Analogously $\mu(\tilde{\beta},2L)$ has been determined
from a Monte Carlo simulation at a nearby value of $\beta$ and then 
applying the appropriate reweighting. Since we use small lattices,
the reweighting technique works very well and it does not increase 
significantly the error bars.

Since we wanted to determine the mass gap with high precision we paid
particular attention to systematic effects. Most of the simulations were 
done on lattices $L\times T$ with pbc and $T = 10L$. 
We computed the correlation function
\be
G(t) = {1\over TL^2} \sum_{x_1,x_2,y} 
   \langle \bsigma_{x_1,y}\cdot \bsigma_{x_2,y+t}\rangle,
\ee
and extracted the mass gap assuming
\be
G(t) \sim \cosh m\left( {T\over2} - t\right)
\ee
for $L \ltapprox t \ltapprox T/2$.
To check for possible systematic deviations we also performed two
simulations with ``cold wall" boundary conditions: in this case 
the spins at one temporal boundary $(t=0)$ where fixed in one direction
$\bsigma^{wall}$ while we used free boundary conditions on the other side.
Of course, in the spatial direction we still used pbc.
In this case the mass gap was extracted from 
\be
G(t) = \sum_x \< \bsigma^{wall}\cdot \bsigma_{t,x}\>
\ee
that is expected to behave as $e^{-m t}$ for $t$ large.
This type of boundary conditions automatically projects out the 
excited modes that have energy 
\begin{eqnarray}
E_l(\beta,L) - E_0 (\beta,L)\, =\, \frac{1}{2\b L}l(l+1)+O(\b ^{-2}),
\end{eqnarray}
and therefore reduces the systematic errors. 
However it has a disadvantage for our purposes: with respect to pbc,
much longer 
simulations are required to obtain the mass gap with the same precision.
We have performed two runs at $L=5,10$ with ``cold wall" boundary conditions
and we have not observed any systematic difference  with respect to the runs 
with pbc (see Table \ref{tabellaMC}). Therefore we believe that our 
results have a systematic error that is smaller than the statistical one.

The results of our simulations are reported in Table \ref{tabellaMC}. 
The total simulation took approximately 5 months of CPU-time on 
a SGI Origin2000. 

\begin{table}
\footnotesize
\begin{tabular}{|c|c|c|c|c|c|c|}
\hline
$L\times T$ & b.c. & $\b_{run}$ & $\tilde{\b}$ & $\mu(\tilde{\b})L$ 
& $N_{stat}$ & $T_{CPU}$(ms)\\
\hline
$5\times 30$   & periodic & $1.212$ & $1.21045\pm 0.00025$ &
 * & $15\cdot 10^6$ & 11/4.5\\
$10\times 100$ & periodic & $1.212$ & * &
$1.2780\pm 0.0019$ & $3.75\cdot 10^6$  & 78/30 \\
\hline
$5\times 50$   & cold wall & $1.212$ & $1.2104\pm 0.0004$ & * &
$84\cdot 10^6$  &  11/8\\
$10\times 60$  & cold wall & $1.212$ & * & $1.278\pm 0.002$ &
$175\cdot 10^6$ &  25/18\\
\hline
$7\times 70$   & periodic  & $1.260$ & $1.2691\pm 0.0004$ & *&
$21.2\cdot 10^6$ &  38/15 \\
$14\times 140$  & periodic & $1.260$ &*& $1.2782\pm 0.0009$ &
$11\cdot 10^6$   & 180/60 
\\
\hline
$10\times 100$  & periodic & $1.340$  & $1.32835\pm 0.00018$ & * &
$16.2\cdot 10^6$  & 78/30 \\
$20\times 200$  & periodic & $1.330$  & * & $1.2738\pm 0.0007$&
$7.2\cdot 10^6$   & 420/122 \\
\hline
\end{tabular}
\caption{Monte Carlo results. $\beta_{run}$ is the value of $\beta$
at which the Monte Carlo simulation was done, 
$\tilde{\beta}$ is the value of $\beta$ such that 
$\mu(\tilde{\beta},L) = 1.0595$. $N_{stat}$ is the number of iterations of
a Wolff algorithm with standard Swendsen-Wang updatings. 
$T_{CPU}$ is the CPU time in ms for a iteration on a SGI Origin2000: the 
first number is the total CPU time spent in our simulation, the second 
number is the time spent in the updating only, i.e. without measuring the 
two-point function.}
\label{tabellaMC}
\end{table}
These results are shown in Fig. \ref{fig3} together with the previously
obtained
ones referring to the {\it standard} \cite{LuschNum} and {\it perfect} 
\cite{HasenPer} actions. Clearly the on-shell action is better than 
the standard action. However its behaviour is worse than that of the 
perfect action which shows no scaling corrections even for $L=5$.
%
%
%
%
%
%
\section{Critical discussion} \label{sec5}
In this paper we have investigated the FSS behaviour 
of the action \reff{ons}. The main point was to understand, in a 
model in which high-statistics data can be generated, how effective
perturbative  improvement is.
Our large-$N$ analysis presented in Sect. \ref{sec3}
shows that tree-level improvement effectively reduces the scaling
corrections. In this limit, simulations with the standard action 
on a lattice of volume $L^2$ would be affected by the same systematic errors
of results obtained on a lattice of size of order $L/2$ using the 
action \reff{ons} (cf. Table \ref{tableFSS1} and Fig. \ref{RNinf}). 
For $N=3$ our numerical data show a similar effect. 
Therefore,
using the tree-level improved action \reff{ons} one can 
perform simulations on a lattice that is four times smaller and still
obtain the same scaling corrections. Of course, in order to make a 
fair comparison of the two actions,
we should take into account the fact that the action \reff{ons} is more
complicated and therefore one Monte Carlo iteration is slower. For instance,
one update with the on-shell action takes 30 ms on a $10\times100$ lattice
(see Table \ref{tabellaMC}), while it takes 11 ms if one uses the 
standard action. Therefore, the time spent in the updating is 
reduced only by 20--30\%. However it is also important to take into account 
the time
spent in the measurement of the observables. In our simulation we measured
the two-point function: the total CPU-time per iteration 
on $10\times100$ lattice is 78 ms, which should be compared to 
300 ms spent by the standard action on a $20\times 200$ lattice.
Using the on-shell action we obtain comparable results for this 
quantity in 1/4 of the 
CPU time.  Notice that we expect 
the two-dimensional $\sigma$-model to be the case in which 
Symanzik improvement has the smaller pay-off. Indeed in this case
the lattice volume increases only as $L^2$, and the algorithm
does not have critical slowing down, so that working with large 
volumes is not so expensive. In QCD the situation would be radically different
since in this case the CPU time to produce an independent configuration 
scales as $L^{4+z}$ with $z\gtapprox 2$. Thus, even a small reduction 
of the needed values of $L$, significantly reduces the computer time. 

Of course, the skeptical reader may think that the better behaviour of 
the improved action is a mere numerical coincidence.
After all, we have simply improved the behaviour of the corrections
to scaling by $\log L$, which is a very slowly varying function: 
therefore, it is not even obvious that, for our range of values of $L$,
say $L\approx 16$--$128$, an improved action behaves better than the
standard one.
Indeed it is possible to write down 
(complicated) actions that are not tree-level improved and yet, for
$L\approx16$--$128$, behave better than the action \reff{ons}, 
and, viceversa, to invent improved actions that, in the same range of 
values of $L$, are worse than the standard action. However the 
relevant question --- unfortunately a not well defined one ---
is whether ``simple" local
tree-level improved actions behave better as long as $L$ is larger than 1. 
In order to understand if there is a positive answer,
we should 
study different actions. We have considered the Symanzik 
tree-level action \reff{sym}. For $N=\infty$ the actions \reff{ons} and 
\reff{sym} behave similarly, and also for $N=3$ no significant 
difference is observed \cite{Alles_unpublished}. Another tree-level
improved action is the perfect action \cite{HasenPer}. Also in this case 
numerical simulations show a better behaviour, see Fig. \ref{fig3}.
Tree-level improvement seems really to work.

Let us now compare the various improved actions in more detail. While the 
theories defined by \reff{ons} and \reff{sym} behave similarly,
the perfect action is vastly better and indeed
it does not show any corrections to scaling (at the level of 1--2\%)
even for $L=5$. A possible explanation of this behaviour 
has been suggested by Hasenfratz and Niedermayer 
\cite{Hasenfratz-Niedermayer_97}. They show that the classically perfect
action used in the simulations reported in Ref. \cite{HasenPer} 
has very small one-loop corrections so that it can be 
effectively considered one-loop improved. As an indication of how 
much one-loop improved an action is, they suggested considering 
the mass gap of states with spatial momentum $p$ in perturbation 
theory at one loop. The general 
analytic calculation of this quantity at one loop is reported in
App. \ref{AppB}. If we define the correlation function
\be
G(t,p) = {1\over L} \sum_{x_1,x_2} 
      \< \s{0,x_1} \cdot \s{t,x_2} \>\, e^{ip\cdot(x_2-x_1)}\; ,
\ee
and the mass gap $\omega(p)$
\be
\omega(p) =\, - \lim_{|t|\to\infty} {\log G(t,p)\over |t|},
\ee
then, we find in perturbation theory
\be
\omega(p)L\, =\, pL + {1\over 2\beta} +\, 
           {1\over L^2} \widehat{\omega}(pL,\beta) + O(\beta^{-2},L^{-4}).
\ee
For the various action we obtain:
\begin{eqnarray}
\mbox{\rm standard} &&\qquad \widehat{\omega}(pL,\beta) =\, 
    - {1\over 12} (pL)^3 + {\pi\over 12 \beta} pL;
\\
\mbox{\rm on-shell} &&\qquad \widehat{\omega}(pL,\beta) =\, 
  -{(pL)^3\over\beta}\left[0.01604 (N-1) + 0.01321\right];
\\
\mbox{\rm Symanzik} &&\qquad \widehat{\omega}(pL,\beta) =\, 
  -{(pL)^3\over\beta}\, 0.01237;
\\
\mbox{\rm perf. $\kappa=2$} &&\qquad \widehat{\omega}(pL,\beta) =\, 
  -{(pL)^3\over\beta}\left[a_{pf} (N-1) - 0.0004\right].
\end{eqnarray}
We have been unable to estimate the constant $a_{pf}$ precisely, but our results
indicate $|a_{pf}|\ltapprox 0.0005$. From the estimates of 
Ref. \cite{Hasenfratz-Niedermayer_97} we would obtain the bound 
$|a_{pf}|\ltapprox 10^{-4}$.

As expected, for tree-level improved actions, there are no $\beta^0$
corrections. Considering now the $O(\beta^{-1})$ terms, we see that
the perfect action has much smaller corrections compared to the 
Symanzik and the on-shell one. Therefore, if the good scaling 
behaviour of the perfect action is related to the fact that it is 
effectively one-loop improved --- in other words, if higher order corrections
play little role --- our results could have been expected on the 
basis of the results for the mass gap we reported above. 
A check of this argument consists in verifying that 
an action that is exactly one-loop improved and is sufficiently local 
has the same good scaling behaviour of the perfect action. A simulation is 
in progress. Of course, at the end of this discussion,
the most important question concerning the effectiveness of 
perfect actions remains unanswered: 
had we to expect {\em a priori} that the perfect action has 
very small one-loop corrections? For $\kappa=+\infty$ the action 
is exactly one-loop improved, but the couplings are long-range so that 
this fact is of no practical interest. The interesting question is what 
happens for those values of $\kappa$ that correspond to sufficiently 
local actions. To answer it we have computed the four-spin
coupling of the perfect action for various values of $\kappa$ and 
correspondingly the correction $\widehat{\omega}(pL,\beta)$.
The results for the perturbative coefficients 
(see Eq. \reff{risultatoOmegaimproved}) 
are shown in Fig. \ref{figmassgap}, where, for each 
value of $\kappa$, we report two points corresponding to two different
truncations of the action, corresponding to including 
respectively 51 and 771 different couplings in the four-spin term
of the action (see App. \ref{AppC} for details).
One immediately sees that the corrections increase strongly 
as $\kappa\to0$, and, for instance for $\kappa = 0.75$, the corrections are 
of the same order of those of the action \reff{ons}. In other words,
the renormalization-group procedure which is the basis of the perfect
action approach does not provide naturally actions that are
``numerically" improved: there are classically
perfect actions that are relatively local and 
that behave no better than \reff{ons}. In conclusion 
we do not see any theoretical reason for the exceptionally good behaviour found 
in Ref. \cite{HasenPer}. Indeed it is not a {\em a priori} obvious 
--- in the sense that it is not built in the approach --- 
that there exists a value of $\kappa$ such that the action is local
and the corrections are small.

%
%
%
%
\section*{Acknowledgments}
The numerical computations presented in this paper have been performed
at the Computer Center of the Scuola Normale Superiore in Pisa.
%
%
%
\appendix

\section{Finite-size scaling in the large-$N$ limit} \label{AppA}
\subsection{Definitions}
In this section we will report some definitions and expressions that will 
be used in our computation of the corrections to FSS.

FSS functions are expressed in terms of the so-called remnant
functions \cite{BF-Arch-Rat-Mech-Anal}. We define 
\begin{eqnarray}
G_{k}(\alpha) &=&  \sum_{n=1}^{\infty} \left[
    (n^2 + \alpha^2)^{k-1/2} - \sum_{m=0}^{k}
     {k-1/2 \choose m} \alpha^{2 m} n^{2 k - 2 m - 1}\right] \; ,
\label{Gkdef}
\\
H_k(\alpha) &=&
     \sum_{n=1}^{\infty} {1\over (n^2+\alpha^2)^{k+1/2}} \; ,
\end{eqnarray}
where $k\ge 0$ (resp. $k\ge 1$) in the first (resp. second) case.
Asymptotic expressions and various properties of the remnant functions
are reported in Ref. \cite{BF-Arch-Rat-Mech-Anal} and in 
Appendix A.1 of Ref. \cite{Caracciolo-Pelissetto_98}.

Our results are expressed in terms of the functions 
$F_0(z;\rho)$ and $F_1(z;\rho)$. They are defined by the asymptotic
expansion of the lattice sum
\be
I_{L,T}(m^2) =\, {1\over LT} \sum_{n_x = 0}^{L-1} \sum_{n_y = 0}^{T-1}
   {1\over \hat{p}^2 + z^2/L^2}
\ee
for $L,T\to\infty$ with $z$ and $\rho=L/T$ fixed. Here
$\hat{p}^2 = 4 \sin^2(p_x/2) + 4 \sin^2(p_y/2)$, 
$p_x = 2 \pi n_x/L$, and $p_y = 2 \pi n_y/T$. 
We obtain
\be
I_{L,T}(m^2) =\, {1\over2\pi} \log L +\,
          F_0(z;\rho) - {z^2\over 16 \pi L^2} \log L +
         {1\over L^2} F_1(z;\rho) +\, O(L^{-4}\log L).
\label{ILTexpansion}
\ee
Explicit expressions for $F_0(z;\rho)$ and $F_1(z;\rho)$ for 
arbitrary $\rho$ are reported in App. B of Ref.
\cite{Caracciolo-Pelissetto_98}. Here we report them only 
for the strip case, i.e. for $\rho = \infty$. We have
\begin{eqnarray}
F_0(z;\infty) &=& {1\over 2z} + {1\over 2 \pi}
      \left(\gamma_E - {1\over2} \log {\pi^2\over2} +
            G_0 \left({z\over2\pi}\right)\right) ,
\label{F0explicito} \\
F_1(z;\infty) &=& {\pi\over6}\left({1\over12} - G_1\left({z\over2\pi}\right)
           \right) - {z\over16} +
           {z^4\over 192\pi^3} H_1\left({z\over2\pi}\right)
\nonumber \\
&& - {z^2\over 16 \pi} \left[ \gamma_E - {1\over2} \log {\pi^2\over2}
     + {2\over3} G_0\left({z\over2\pi}\right) - {1\over3}\right]\; .
\label{F1explicito}
\end{eqnarray}
We will also need the asymptotic expansion for $z\to0$. We have
\be
F_0(z;\infty) = {1\over 2z} + \overline{F}_{00} + O(z^2)
\label{F0smallzstrip} 
\ee
with
\be
\overline{F}_{00} = {1\over 2\pi}
    \left (\gamma_E - \log \pi + {1\over2}\log 2 \right ),
\label{Fbar00}
\ee
where $\gamma_E \approx 0.577215664902$ is Euler constant.

\subsection{Computation of the FSS functions}

In this Appendix we will compute the FSS functions and their leading
corrections in the large-$N$ limit for the theory defined
by the action \reff{azione-generale}. We will follow
closely the strategy of Ref. \cite{Caracciolo-Pelissetto_98} to which 
we refer for the derivation of the basic results.

We will consider a square lattice of size $L\times T$. The FSS limit
corresponds to $\beta\to\infty$, $L,T\to \infty$ with 
$L/T\equiv \rho$ and $m_{L,T} L\equiv z$ fixed. 

The first step is the derivation of the asymptotic expression 
for $\omega_{L,T}$. From the gap equations 
\reff{gap1} and \reff{gap2} we obtain
\begin{equation}
-{1\over 2\alpha_3} \omega_{L,T} \ =\,
{  {\D 1\over \D LT} {\D\sum_p} 
   {\D \hat{K}(p)\over \D \hat{w}(p;\omega_{L,T})+m^2_{L,T}}
\over 
   {\D 1\over \D LT} {\D\sum_p} {\D 1\over \D \hat{w}(p;\omega_{L,T})+m^2_{L,T}}
}\; .
\end{equation}
{}From this equation we immediately see that, for $m_{L,T}\to 0$, the 
r.h.s. goes to zero for any $L,T$, including $L,T=\infty$. Therefore 
in the FSS limit $\omega_{L,T}\to 0$. Now, at least 
for $\omega_{L,T}$ sufficiently
small, $\hat{w}(p;\omega_{L,T})\ge 0$ everywhere since $w(p)\ge 0$; 
moreover $\hat{w}(p;\omega_{L,T})$ vanishes only for $p=0$ in the Brillouin 
zone.  Therefore, we can use the results of Ref.
\cite{Caracciolo-Pelissetto_98} and we can write 
\begin{equation}
\omega_{L,T} \ =\,
{ -2 \alpha_3 \Xi_1(\omega_{L,T}) \over 
  {1\over 2\pi} \log L + F_0(z;\rho) + \Lambda_0 (\omega_{L,T})}
+ O\left({\log L\over L^2}\right),
\label{omegaLT}
\end{equation}
where
\begin{eqnarray}
\Lambda_0(\omega)&=&\int {d^2p\over (2\pi)^2}
\left( \frac{1}{\hat{w}(p;\omega)}-\frac{1}{\hat{p}^2}\right),
\\
\Xi_1(\omega)&=&\int {d^2p\over (2\pi)^2}
      \frac{\hat{K}(p)}{\hat{w}(p;\omega)},
\end{eqnarray}
and $F_0(z;\rho)$ is defined in Eq. \reff{F0explicito}.

Eq. \reff{omegaLT} shows that $\omega_{L,T}$ goes to zero 
in the FSS limit logarithmically and that it admits an expansion in
powers of $1/\log L$. Explicitly we have 
\be
\omega_{L,T} = - {4 \pi \alpha_3 \Xi_1(0)\over \log L} 
\left[1 + {2 \pi \over \log L} \left(-2 \alpha_3 \Xi_1(0) 
    + 2 \alpha_3 \Xi_2(0) - F_0(z;\rho) - \Lambda_0(0)\right)\right] +
   O(\log^{-3}L),
\label{omegaLTespansa}
\ee
where
\be
\Xi_2(\omega) =\ \int {d^2p\over (2\pi)^2}
      \frac{\hat{K}(p)^2}{\hat{w}(p;\omega)^2}.
\ee
Eq. \reff{omegaLT} defines $\omega_{L,T}$ implicitly as a function 
of $z$, $L$, and $\rho$, and it should be solved exactly if we want
to obtain the corrections to FSS up to terms of order
$\log^2 L/L^4$. In the following we will obtain the corrections to FSS 
parametrized in terms of $z$ and $\omega_{L,T}$; it is understood
that $\omega_{L,T}$, for each $z,L$ and $\rho$, is the solution of 
Eq. \reff{omegaLT}.

Let us now go back discussing the validity of Eq. \reff{omegaLT}. 
As already mentioned above, it is valid only for those values of 
$\omega_{L,T}$ such that $\hat{w}(p;\omega)\ge 0$. A second limitation
derives from the asymptotic expansion of the denominator in the 
r.h.s. Indeed, if $\hat{w}(p;\omega) \ge 0$, the sum 
$\sum_p (\hat{w}(p;\omega) + m^2_{L,T})^{-1}$ is always positive.
On the other hand the asymptotic expansion is negative for 
large values of $z$. In particular, for each $L$ and $\omega_{L,T}$
there is a unique value $z_c(L,\omega_{L,T})$ such that 
\be
{1\over 2\pi} \log L + F_0(z_c(L,\omega_{L,T});\rho) + 
 \Lambda_0(\omega_{L,T}) = 0.
\ee
It is easy to check, using the results of
Ref. \cite{Caracciolo-Pelissetto_98}, that the asymptotic expansion
makes sense only for $z \ll z_c (L,\omega_{L,T})$ and that 
$z_c(L,\omega_{L,T})$ behaves as $L$ for $L\to\infty$. In other words
the asymptotic expansion of the lattice sum in powers of $1/L^2$ is 
not uniform in $z$:
the error increases as $z\to\infty$. This 
fact is not unexpected. Indeed since for $L\to\infty$
we find $z_c(L,\omega_{L,T}) \sim L$, our discussion tells
us that the expansion is valid only for $L\gg z$, i.e.
for $m_{L,T}\ll 1$, that is when the correlation length
is much larger than the lattice spacing. In Fig. \ref{figomega} we report
$\omega_{L,T}$ for two different values of $L$ and $\rho =\infty$. 
The non-uniformity is clearly evident: the expansion worsens as $z\to\infty$
at $L$ fixed.

A second important fact that is evident in Eq. \reff{omegaLT} is that
the perturbative (PT) limit $z\to 0$ at fixed $L$ followed by the limit 
$L\to\infty$ does not commute with the FSS limit. 

Let us consider for simplicity the strip case $\rho=\infty$ and use the 
asymptotic expansion \reff{F0smallzstrip}.
Then, in the perturbative limit,
we obtain
\be
\omega_{L,\infty} = -4 \alpha_3 \Xi_1(0) z
 \left[ 1 - {z\over \pi} \left(\log L + c_\omega\right)\right] 
  + O(z^3 \log^2 L),
\ee
with
\be
c_\omega = 4 \pi \alpha_3 \left[\Xi_1(0) - \Xi_2(0)\right] + 
    2 \pi \overline{F}_{00} + 2 \pi \Lambda_0(0).
\ee
In the PT limit $\omega_{L,\infty}$ has an expansion in powers of 
$z$ such that the coefficient of $z^n$ is of order $(\log L)^{n-1}$. 
This expansion is clearly different from the expansion 
\reff{omegaLTespansa}, that, for $z\to 0$, becomes:
\be
\omega_{L,T} = {-4 \pi \alpha_3 \Xi_1(0)\over \log L} 
  \left[1 - {\pi\over z\log L} (1 + O(z)) + O(\log^{-2} L)\right].
\ee
In the FSS limit we have a different expansion in powers of $1/\log L$
with coefficients that diverge for $z\to 0$.

The next step in the derivation consists in obtaining the relation between 
$\omega_{L,T}$ and $\omega_\infty$. Starting from the second gap equation
\reff{gap2} we have
\be
{(1 + \omega_{L,T})\omega_{L,T} \over 
 (1 + \omega_\infty)\omega_\infty} 
  \int {d^2p\over (2\pi)^2}
   {\hat{K}(p)\over \hat{w}(p;\omega_\infty)+m^2_\infty} =\, 
  {1\over LT} \sum_p {\hat{K}(p)\over \hat{w}(p;\omega_{L,T})+m^2_{L,T}}.
\label{A.19}
\ee
In the FSS limit, assuming always $\omega_{L,T}$ and 
$\omega_\infty$ small enough so that $\hat{w}(p;\omega) \ge 0$,
using the results of Ref. \cite{Caracciolo-Pelissetto_98},
it is easy to compute 
\begin{eqnarray}
\hskip -27pt
{1\over LT} \sum_p {\hat{K}(p)\over \hat{w}(p;\omega_{L,T})+m^2_{L,T}} &=&
   \Xi_1(\omega_{L,T}) - 
    m^2_{L,T} \left({1\over 2\pi} \log L + F_0(z;\rho) + 
    \Xi_3(\omega_{L,T})\right)\! ,
\\
\hskip -27pt
\int {d^2p\over (2\pi)^2} 
   {\hat{K}(p)\over \hat{w}(p;\omega_\infty)+m^2_\infty} &=&
  \Xi_1(\omega_\infty)  - m^2_\infty
   \left(- {1\over 4\pi} \log {m^2_\infty\over 32} +
    \Xi_3(\omega_\infty)\right) ,
\end{eqnarray}
with corrections of order $m^4_{L,T}\log m_{L,T}$ and 
$m^4_\infty\log m_\infty$ respectively. 
Here
\be
\Xi_3(\omega) =\, 
   \int {d^2p\over (2\pi)^2} 
    \left({\hat{K}(p)\over \hat{w}(p;\omega)^2} - {1\over \hat{p}^2}
    \right).
\ee
Using Eq. \reff{A.19} we obtain
\be
\omega_\infty = \omega_{L,T} \left (1 + 
    \Delta_\omega(z,\omega_{L,T};\rho;L)
     {1\over L^2} + O(\log^{2} L/L^4)\right),
\label{omegainfty}
\ee
where 
\begin{eqnarray}
\lefteqn{\hspace{-1cm} \Delta_\omega(z,\omega;\rho;L) = }\nonumber \\
&&\hspace{-1cm} 
 {1+\omega \over (1 + \omega) \Xi_1(\omega) + \omega \Xi_2(\omega)} 
 \left(z^2 - 32 e^{-4 \pi F_0(z;\rho)} \right) 
 \left({1\over 2\pi} \log L + F_0(z;\rho) + \Xi_3(\omega)\right).
\end{eqnarray}
In the calculation of this expression we have already used 
the FSS result for $m^2_\infty/m^2_{L,T}$ that will be derived in
the following. The function $\Delta_\omega(z,\omega;\rho;L)$ scales 
as $\log L$ for $L\to\infty$ and it has an expansion 
in powers of $1/\log L$. Explicitly
\be
\Delta_\omega(z,\omega_{L,T};\rho;L) = 
   \sum_{k=-1}^\infty {\delta_{\omega,k}(z;\rho)\over \log^k L}.
\ee
The first coefficients are 
\begin{eqnarray} 
\delta_{\omega,-1} (z;\rho) &=& {1\over 2 \pi \Xi_1(0)}
   \left(z^2 - 32 e^{-4 \pi F_0(z;\rho)} \right) ,
\\
\delta_{\omega,0} (z;\rho) &= &
{1\over \Xi_1(0)} \left(z^2 - 32 e^{-4 \pi F_0(z;\rho)} \right)
 \left[2 \alpha_3 \Xi_1(0) + F_0(z;\rho) + \Xi_3(0)\right].
\end{eqnarray}
Finally let us compute the corrections to FSS for the ratio 
$m^2_\infty/ m^2_{L,T}$. Using Eq. \reff{gap1}
we obtain
\be
{1\over 1 + \omega_\infty} 
\int {d^2p\over (2\pi)^2}
   {1 \over \hat{w}(p;\omega_\infty)+m^2_\infty} = 
{1\over 1 + \omega_{L,T}}
{1\over LT} \sum_p {1 \over \hat{w}(p;\omega_{L,T})+m^2_{L,T}}.
\ee
Using the results of Ref. \cite{Caracciolo-Pelissetto_98} and Eq.
\reff{omegainfty}, we obtain
\be
{m^2_\infty\over m^2_{L,T}} = 
  f_m(z;\rho)\left(1 + \Delta_m(z,\omega_{L,T};\rho;L)
     {1\over L^2} + O(\log^{2} L/L^4)\right)
\label{espansioneasminfsumLT}
\ee
where
\be
f_m(z;\rho) = {32\over z^2} e^{-4 \pi F_0(z;\rho)},
\ee
and 
\begin{eqnarray}
\Delta_m(z,\omega_{L,T};\rho;L) &=&
        {1\over4} (12 \hat{\alpha}_1 + 16 \hat{\alpha}_2 -1)
        \left( 32 e^{-4 \pi F_0(z;\rho)} - z^2\right) \log L + 
\nonumber \\
&& \hskip -35pt
   16\pi (12 \hat{\alpha}_1 + 16 \hat{\alpha}_2 - 1) F_0(z;\rho)
         e^{-4 \pi F_0(z;\rho)}
\nonumber \\ [2mm]
&& \hskip -35pt 
    - 4 (8\hat{\alpha}_1 + 8\hat{\alpha}_2 - 1) e^{-4 \pi F_0(z;\rho)}
    - {4 \pi}\left( {\cal F}_1(z,\omega_{L,T};\rho) +
      32 e^{-4 \pi F_0(z;\rho)} \Lambda_1(\omega_{L,T})\right)
\nonumber \\ [2mm]
&& \hskip -35pt
   - {4 \pi \omega_{L,T}\over 1 + \omega_{L,T}} 
   \left({1\over 2\pi} \log L + F_0(z;\rho) + \Xi_3(\omega_{L,T})\right)
   \Delta_\omega(z,\omega_{L,T};\rho;L).
\label{Deltam}
\end{eqnarray}
Here
\begin{eqnarray}
{\cal F}_1(z,\omega;\rho) &=& (1-12 \hat{\alpha}_1) F_1(z;\rho) +
     z^2 \left({\hat{\alpha}_1\over 8\pi} - \Lambda_1(\omega)\right)
    + 2 \hat{\alpha}_2 z^2 F_0(z;\rho) +
     {\hat{\alpha}_2\over2} z^3 {\partial F_0\over \partial z} (z;\rho) ,
 \nonumber \\ [-2mm]
{}
\label{F1storto} \\
\Lambda_1(\omega) &=& \int {d^2p\over (2\pi)^2}
     \left( {1\over \hat{w}(p;\omega)^2} - {1\over (\hat{p}^2)^2} +
     {2\over (\hat{p}^2)^3} \left( \hat{\alpha}_1 \sum_\mu \hat{p}^4_\mu
       + \hat{\alpha}_2 (\hat{p}^2)^2 \right) \right). 
\label{Lambda1}
\end{eqnarray}
The variables $\hat{\alpha}_1$ and $\hat{\alpha}_2$  are functions of 
$\omega$ defined by the asymptotic expansion for $p\to 0$
\be
\hat{w}(p;\omega)   = \,  \p^2 + \a{1}\sum_{\mu}\p_{\mu}^4
             +\a{2} (\p^2)^2 + O(p^6).
\ee
The function $\Delta_m(z,\omega;\rho;L)$ scales 
as $\log L$ for $L\to\infty$ and it has an expansion 
in powers of $1/\log L$. Explicitly
\be
\Delta_m(z,\omega_{L,T};\rho;L) = 
   \sum_{k=-1}^\infty {\delta_{m,k}(z;\rho)\over \log^k L}.
\ee
The first coefficients are 
\begin{eqnarray} 
\delta_{m,-1} (z;\rho) &=& 
    - {1\over4} (12 \alpha_1 + 16 \alpha_2 - 16 \alpha_3 -1)
   \left(z^2 - 32 e^{-4 \pi F_0(z;\rho)} \right) 
\\
\delta_{m,0} (z;\rho) &= & - {\pi^2\over 18} (1 - 12 \alpha_1) + 
  {\pi z\over 4} \left(1 - 12 \alpha_1 - 12 \alpha_2 + 16 \alpha_3\right)
  + O(z^2).
\end{eqnarray}
One immediately sees that if the action is tree-level improved 
$\delta_{m,-1} (z;\rho) = 0$. The next coefficient 
$\delta_{m,0} (z;\rho)$ instead does not vanish in general.
In Fig. \ref{Deltainf} we report the function $\Delta_m(z,\omega_{L,T};\rho;L)$
for $\rho=\infty$ together with the exact deviations from finite-size
scaling. Notice that for $L=10$ the higher-order terms in $1/L$ 
behaving as $\log L/L^4$ still play an important role for 
$z\approx2-10$, while they are negligible for $L=30$.

Since our class of actions for $\alpha_3\not=0$ is only on-shell 
improved, we expect to find an improved behaviour only in on-shell
quantities. On the strip ($\rho=\infty$) let us determine the 
FSS behaviour of the mass gap $\mu(L)$. In the FSS limit with 
$L \mu(L)\equiv x$ fixed we have
\be
{\mu(\infty)^2\over \mu(L)^2} = f_\mu(x) 
  \left(1 + \Delta_\mu(x,\omega_{L,\infty};L) {1\over L^2} + 
   O(L^{-4}\log^2 L ) \right)
\ee
where $f_\mu(x) = f_m(x)$ and 
\begin{eqnarray}
\Delta_{\mu}(x,\omega;L) &=& 
\Delta_{m}(x,\omega;\infty;L)+\frac{8}{3}(12\ah_1 +
          12\ah_2 -1)e^{-4\pi F_0(x;\infty)} +
\nonumber\\
&&   +(12\ah_1 +12\ah_2 -1)\frac{\pi x^3}{6}
       \frac{\partial F_0(x;\infty)}{\partial x}
\end{eqnarray}
For $L\to\infty$, $\Delta_{\mu}(x,\omega;L)$ has an expansion in powers
of $1/\log L$:
\be
\Delta_\mu(x,\omega_{L,T};L) = 
   \sum_{k=-1}^\infty {\delta_{\mu,k}(x)\over \log^k L}.
\ee
It is easy to see that $\delta_{\mu,-1} (x) = \delta_{m,-1} (x;\infty)$.
Therefore the leading term vanishes for tree-level improved actions.
If $\alpha_1 = {1\over 12}$ and $\alpha_2 = \alpha_3$ we have
\begin{eqnarray}
&& \hskip -1truecm 
\delta_{\mu,0}(x) =\,
    \left\{ 8\pi \alpha_2\left[\Lambda_0(0)-2\alpha_2\,
           \Xi_1(0)-2\,\Xi_2(0)-2\,\Xi_3(0)\right]
     -4\pi\left[\Lambda_1(0)-\frac{1}{96\pi}\right]
\right.  \nonumber \\
&& \left.
   -4\pi\alpha_2\left[3\left(\beta_1-\frac{1}{12}\right)
   +4\left(\beta_2-\alpha_2\right)\right]\Xi_1(0)
   \right\} \left(32 e^{-4\pi F_0(x;\rho)} -x^2\right),
\label{eqA41}
\\
&& \hskip -1truecm
\delta_{\mu,1}(x) =\,
       -4\pi\alpha_3\Xi_1(0)\left[\frac{2\pi}{3}
        \left(\beta_1-\frac{1}{12}\right)-x\left(
        \beta_1-\frac{1}{12}-32\pi^2\alpha_3\right)+ O(x^2)\right],
\label{deltamu1}
\end{eqnarray}
where $\beta_1$ and $\beta_2$ are defined from 
\be
K(p) = \hat{p}^2 + \beta_1 \hat{p}^4 + \beta_2 (\hat{p}^2)^2 + O(p^6).
\label{Kexpan}
\ee
{}From Eq. \reff{eqA41} we immediately see that it is possible to choose $w(p)$
and $K(p)$ in such a way that $\delta_{\mu,0}(x) = 0$. Actions 
satisfying this condition are one-loop on-shell improved.
On the other hand, from Eq. (\ref{deltamu1}), one immediately convinces 
oneself that $\delta_{\mu,1}(x)$ never vanishes unless $\alpha_3=0$.
In other words, actions of the form (\ref{azione-generale}) with
$\alpha_3\not=0$ cannot be two-loop improved. This is not unexpected since 
six-spin couplings are needed for two-loop improvement.

\section{Perturbative computation of the mass gap for spatial momentum
$p\not=0$} \label{AppB}

In this Appendix we will compute the mass gap for states with spatial 
momentum $p$ for the most general action with two-spin and four-spin
couplings:
\be
S(\sg) = \frac{1}{2}\sum_{x,y}w_{xy}\sg_x\cdot\sg_y +
\sum_{x_1 x_2 x_3 x_4}
c_{x_1 x_2; x_3 x_4}(\sg_{x_1}\cdot\sg_{x_2}-1)(\sg_{x_3}\cdot\sg_{x_4}-1)\; .
\label{ac}
\ee
Notice that 
the couplings $c_{x_1 x_2; x_3 x_4}$ are well defined only for 
$x_1\not= x_2$ and $x_3\not= x_4$. 
This action reduces to \reff{azione-generale} if
\be
c_{x_1 x_2;x_3 x_4} = -\frac{\alpha_3}{8}K_{x_1 x_2}K_{x_3 x_4}
  (\delta_{x_1 x_3}+\delta_{x_2 x_3}+\delta_{x_1 x_4}+\delta_{x_2 x_4})\; .
\label{cKK}
\ee
We introduce the Fourier transform
\be
c(p;q,k) \equiv \sum_{x_2 x_3 x_4}\, c_{x_1 x_2; x_3 x_4}\,
    e^{i\frac{p}{2}(x_1+x_2-x_3-x_4)}e^{iq(x_1-x_2)}e^{ik(x_3-x_4)},
\ee
and define
\begin{eqnarray}
C(p;q,k)&\equiv& c(p;q,k)-c(p;p/2,k)-c(p;q,p/2)+c(p;p/2,p/2), \\ [2mm]
f(p,q)  &\equiv& C\left(p+q;\frac{p-q}{2},\frac{p-q}{2}\right).
\end{eqnarray}
For $p,q,k\to 0$, keeping into account the lattice symmetries,
we have
\begin{eqnarray}
C(p;q,k) &= &\frac{1}{16}(\alpha+\beta)(p^2)^2+\frac{1}{16}\gamma\sum_{\mu}
p_{\mu}^4-\frac{1}{4}\alpha p^2(q^2+k^2)-\nonumber\\
&&-\frac{1}{4}\beta
\left[(p\cdot q)^2+(p\cdot k)^2\right]-\frac{1}{4}\gamma\sum_{\mu}
(p^2_{\mu}q^2_{\mu}+p^2_{\mu}k^2_{\mu})+\nonumber\\
&&+\alpha q^2 k^2 +\beta (q\cdot k)^2+\gamma \sum_{\mu}q^2_{\mu}k^2_{\mu}
+O(p^6,p^4 q^2,\ldots) \\
f(p,q) &= & \alpha (p\cdot q)^2 + 
            {\beta\over2} \left[ (p\cdot q)^2 + p^2 q^2\right] +
            \gamma \sum_\mu p^2_\mu q^2_\mu +O(p^6,p^4 q^2,\ldots),
\end{eqnarray}
where $\alpha$, $\beta$ and $\gamma$ are free parameters.
If $c_{x_1 x_2;x_3 x_4}$ is given by Eq. \reff{cKK}, then
$\beta=\gamma=0$ and $\alpha=-\alpha_3/2$.
The general conditions that make the action \reff{ac} tree-level
on-shell improved have been discussed in Ref. \cite{Hasenfratz-Niedermayer_97}.
The two-point function is improved if $\alpha_1 = 1/12$, while improvement 
of the four-point function gives
\be
\alpha = - {1\over2} \alpha_2, \qquad \beta=\, \gamma =\, 0.
\ee
Let us now compute the mass gap. Explicitly we consider 
\be
G(t;\overline{p}) = 
  {1\over L}\sum_{x_1,x_2} \< \bsigma_{0,x_1} \cdot \bsigma_{t,x_2}\>
   e^{i \overline{p} (x_2 - x_1)}
\ee
on a strip $\infty\times L$ with periodic boundary conditions in the 
spatial direction. For large $|t|$ we have 
\be
G(t;\overline{p}) \sim e^{-|t| \omega(\overline{p})}.
\ee
The calculation at one loop for $\overline{p}=0$ is reported in Ref. 
\cite{RV,Farchioni-etal}. 
Here we will repeat the calculation for $\overline{p}\not=0$. 
At the order we are interested in we expand
\be
\omega(\overline{p}) = \omega_0(\overline{p}) + 
          {1\over \beta} \omega_1(\overline{p}) + O(\beta^{-2}).
\ee
The tree-level term $\omega_0(\overline{p})$ is easily computed. 
Indeed at tree level we obtain for $\overline{p}\not=0$
\be
G(t;\overline{p}) = {N-1\over\beta} 
   \int^\pi_{-\pi} {dq\over 2\pi} {e^{i q t}\over w(q,\overline{p})}.
\ee
$w(q,\overline{p})$ is a sum of trigonometric functions and therefore 
it is analytic in the whole complex $q$-plane. It is then easy 
to compute the integral using Cauchy's theorem. For $|t|\to\infty$ 
we have
\be
G(t;\overline{p}) =
    {N-1\over\beta} {e^{-\omega_0(\overline{p}) |t|} \over D(\overline{p})},
\ee
where $i \omega_0(\overline{p})$  is the zero of $w(q,\overline{p})$ 
with the smallest positive imaginary part\footnote{We assume here that 
$\omega_0(\overline{p})$ is real. This is not generically true, although it 
is always verified for $\overline{p}\to 0$.} and 
\be
D(\overline{p}) = - i \left( {\partial w(q,\overline{p})\over \partial q}
    \right)_{q = i \omega_0(\overline{p})}.
\ee
For $\overline{p}\to 0$ we have
\begin{eqnarray}
\omega_0(\m) &=& \m+ \left(\alpha_1-\frac{1}{12}\right)\m^3+O(\m^5),
\\
D(\m)        &=& 
     2\m \left[ 1-\left(\alpha_1-\frac{1}{12} \right) \m^2 + O(\m^4) \right] .
\end{eqnarray}
For tree-level on-shell improved actions we have $\alpha_1 = 1/12$. Therefore
for this class of actions $O(a^2)$ corrections vanish as expected.

Let us now consider the one-loop correction. In this case one should pay
special attention to the boundary conditions in the temporal direction.
As usual we consider a lattice of size $L\times T$ with free 
boundary conditions in the temporal direction and we will use the 
explicit expression for the two-point
function reported in \cite{RV}. $G(t;\overline{p})$ is then obtained 
taking the limit $T\to \infty$.
For $\overline{p}\not=0$ we obtain the final expression
\begin{eqnarray} 
  \omega_1(\overline{p} L)L & = & 
       \frac{1}{2}+\frac{L}{D(\m)}\left[\frac{1}{L}
       \sum_{q_1}\int\frac{dq_0}{2\pi}\frac{w(p+q)-w(p)}{w(q)}-1\right]+
\nonumber \\
&& + 4(N-1) \frac{L}{D(\m)}\left[\frac{1}{L}
    \sum_{q_1}\int\frac{dq_0}{2\pi}\frac{1}{w(q)} C(0;p,q) \right] +
\nonumber  \\
&& + 8\frac{L}{D(\m)}\left[\frac{1}{L}
\sum_{q_1}\int\frac{dq_0}{2\pi}\frac{1}{w(q)}\left(f(p,q)-f(p,0)\right)\right],
\label{esa}
\end{eqnarray}
where in the r.h.s. $p\equiv(i\omega_0(\m),\m)$.
We can then expand Eq. \reff{esa} in powers of $1/L$ for $\overline{p} \to 0$
with  $\m L$ fixed. We obtain
\be
\omega_1 (\m)L  =  \frac{1}{2}+\frac{\Omega(\m L)}{L^2}+O(L^{-4}),
\ee
where 
\begin{eqnarray}
\Omega(\m L) &=& \frac{\pi}{12}\m L\left(1-12\alpha_1-8\alpha_2
-16\alpha-16\gamma-8\beta\right)-\frac{2\pi}{3}(N-1)\m L (\beta+\gamma )-
\nonumber\\
&&-\left(\alpha_1 - \frac{1}{12}\right)(N-1)(\m L)^3 E+
(N-1)(\m L)^3 D+\nonumber\\
&&+ (\m L)^3\left[A-3B-F-
2\left(\alpha_1 - \frac{1}{12}\right)\left(G-\frac{1}{4}C\right)\right].
\label{risultatoOmega}
\end{eqnarray}
The constants $A,B,C,D,E,F$, and $G$ are explicitly given by
\begin{eqnarray}
A&\equiv&\frac{1}{48}\int\frac{d^2q}{(2\pi)^2}\frac{1}{w(q)}\sum_{\mu}\left[
\frac{\partial^4 w}{\partial q_{\mu}^4}(q)-
\frac{\partial^4 w}{\partial q_{\mu}^4}(0)\right], \\
B&\equiv&\frac{1}{24}\int\frac{d^2q}{(2\pi)^2}\frac{1}{w(q)}\left[
\frac{\partial^4 w}{\partial q_0^2\partial q_1^2}(q)-
\frac{\partial^4 w}{\partial q_0^2\partial q_1^2}(0)\right],\\
C&\equiv&\frac{1}{2}\int\frac{d^2q}{(2\pi)^2}\frac{1}{w(q)}\sum_{\mu}\left[
\frac{\partial^2 w}{\partial q_{\mu}^2}(q)-
\frac{\partial^2 w}{\partial q_{\mu}^2}(0)\right],\\
D&\equiv&\frac{1}{6}\int\frac{d^2q}{(2\pi)^2}\frac{1}{w(q)}
\left[\left.\frac{\partial^4C}{\partial p_0^4}(0;p,q)\right|_{p=0}
-3\left.\frac{\partial^4C}{\partial p_0^2\partial p_1^2}(0;p,q)\right|_{p=0}
\right],\\
E&\equiv&\int\frac{d^2q}{(2\pi)^2}\frac{1}{w(q)}
\left[\left.\frac{\partial^2C}{\partial p_0^2}(0;p,q)\right|_{p=0}
+ \left.\frac{\partial^2C}{\partial p_1^2}(0;p,q)\right|_{p=0}
\right],\\
F &\equiv &-\frac{1}{3}\int\frac{d^2q}{(2\pi)^2}\frac{1}{w(q)}
\left[\left.\frac{\partial^4f}{\partial p_0^4}(p,q)\right|_{p=0}
-3\left.\frac{\partial^4f}{\partial p_0^2\partial p_1^2}(p,q)\right|_{p=0}
\right], \\
G&\equiv&\int\frac{d^2q}{(2\pi)^2}\frac{1}{w(q)}
\left[\left.\frac{\partial^2f}{\partial p_0^2}(p,q)\right|_{p=0}
+ \left.\frac{\partial^2f}{\partial p_1^2}(p,q)\right|_{p=0}
\right].
\end{eqnarray}
For the action \reff{azione-generale}, these expressions simplify becoming
\begin{eqnarray}
D &=& -2 \alpha_3 \left(\beta_1 - {1\over12}\right)\,
         \int\frac{d^2q}{(2\pi)^2}\frac{K(q)}{w(q)},
\\
E &=& - 2\alpha_3 \int\frac{d^2q}{(2\pi)^2}\frac{K(q)}{w(q)},
\\
F &=& - {3\alpha_3\over2}\, \int\frac{d^2q}{(2\pi)^2}\frac{1}{w(q)}
   \left\{ {1\over 4} \left[
     \frac{\partial^2K}{\partial q_0^2}(q)-\frac{\partial^2K}{\partial q_1^2}
     (q)\right]^2 -\left[
     \frac{\partial^2 K}{\partial q_0\partial q_1}\right]^2 
    \right. \nonumber \\
&& \left. 
+\frac{1}{3}
\sum_{\mu}\frac{\partial K}{\partial q_{\mu}}
\frac{\partial^3 K}{\partial q_{\mu}^3}   -
\frac{\partial K}{\partial q_0}
\frac{\partial^3 K}{\partial q_0\partial q_1^2} -
\frac{\partial K}{\partial q_1}
\frac{\partial^3 K}{\partial q_1\partial q_0^2}\right\} ,
\\
G &=& - {\alpha_3\over4} \int\frac{d^2q}{(2\pi)^2}\frac{1}{w(q)}\left[ 
        \left(\frac{\partial K}{\partial q_0}(q)\right)^2+
        \left(\frac{\partial K}{\partial q_1}(q)\right)^2\right],
\end{eqnarray}
with $\beta_1$ defined in Eq. \reff{Kexpan}. 

If the action is tree-level on-shell improved Eq. \reff{risultatoOmega} 
reduces to 
\be
\Omega(\overline{p}L)\, =\, 
 (\overline{p}L)^3\left[ (N-1) D + A - 3 B - F\right].
\label{risultatoOmegaimproved}
\ee
Notice that if we require the action \reff{risultatoOmega} to be
tree-level improved to order $O(a^4)$ we obtain the additional condition
$\beta_1 = 1/12$. In this case we also have $D=0$.

Computing numerically the various integrals,
we obtain for the various actions we have introduced in the text:
\begin{enumerate}
\item Standard action \reff{std}
\be
\Omega(\m L) = {\pi \over 12} \m L ;
\ee
\item Symanzik action \reff{sym}
\be
\Omega(\m L) = - 0.01237 ( \m L)^3 ;
\ee
\item Diagonal action \reff{dia}
\be
\Omega(\m L) = {\pi \over 18} \m L  - 0.00315 ( \m L)^3 ;
\ee
\item On-shell action \reff{ons}
\be
\Omega(\m L) = - 0.01321 ( \m L)^3 - 0.01604 (N-1) ( \m L)^3.
\ee
\end{enumerate}
In particular for $N=3$ and $\m L = 2\pi$ we obtain in the four cases: 
$\Omega(2\pi) = 1.64493$, $-3.0698$, $0.3163$, $-11.235$. 
For the first two cases we can compare with the 
numerical results of Ref. \cite{Hasenfratz-Niedermayer_97} finding good 
agreement.

We have also computed $\Omega(\m L)$ for the one-loop 
Symanzik action of Ref. \cite{Symanzik}. In this case we should add
to Eq. \reff{risultatoOmega} the tree-level contribution of the 
$1/\beta$ corrections appearing in the action. As expected the sum of the 
two terms vanishes. We should note that this cancellation 
does not happen for the Symanzik action that has been used in the
simulations \cite{Montvay} and that is therefore clearly incorrect.

\section{Mass gap for the perfect action} \label{AppC}

In this Appendix we want to compute the mass gap for $\overline{p}\not=0$
at one-loop for a generic perfect action, generalizing the results
of Ref. \cite{Hasenfratz-Niedermayer_97}. The first step consists in 
deriving the perfect action for $\kappa \not=2$. For our calculation 
we will be only interested in the two-spin and in the four-spin couplings.
We will therefore consider the generic model \reff{ac}.
With the standard parametrization $\sg_x = (\ph_x,\sqrt{1-\ph_x^2})$
we obtain
\be
S(\ph)  =\,  \frac{1}{2}\sum_{x,y}w_{xy}\ph_x\cdot\ph_y +
    \sum_{x_1 x_2 x_3 x_4}
    \overline{c}_{x_1 x_2; x_3 x_4}
   (\ph_{x_1}\cdot\ph_{x_2}) (\ph_{x_3}\cdot\ph_{x_4})+\dots
\label{actionintermsphi}
\ee
where
\begin{eqnarray}
\overline{c}_{x_1 x_2; x_3 x_4} & = & c_{x_1 x_2; x_3 x_4}-
     \delta_{x_1 x_2}\sum_z c_{x_1 z; x_3 x_4} -
     \delta_{x_3 x_4}\sum_z c_{x_1 x_2; x_3 z} +
\nonumber\\
&&  +\delta_{x_1 x_2}\delta_{x_3 x_4}\sum_{z z'} c_{x_1 z; x_3 z'}+
    \frac{1}{8}w_{x_1 x_3}\delta_{x_1 x_2}\delta_{x_3 x_4}.
\label{cbar}
\end{eqnarray}
It is easy to see that $\overline{c}_{x_1 x_2; x_3 x_4}$ satisfies the
constraint
\be
\sum_{z} 
      \overline{c}_{x_1 z; x_2 x_3}  =\,
      {1\over8} w_{x_1 x_2} \delta_{x_2 x_3}, 
\label{C4}
\ee
that is related to the $O(N)$-invariance of the theory. In other words,
an action of the form \reff{actionintermsphi} is $O(N)$-invariant
up to terms of order $\phi^6$ if and only if Eq.
\reff{C4} is satisfied.

The couplings $w_{xy}$ and $\overline{c}_{x_1 x_2; x_3 x_4}$ are obtained 
requiring the action to be a fixed point of a family of 
renormalization-group (RG) transformations
\cite{HasenPer} labelled by a parameter $\kappa$. 
One finds that $w_{xy}$ is the perfect laplacian
defined in Eq. \reff{perfect-laplacian}. The four-spin coupling 
is the fixed point of the equation
\be
\overline{c}'_{z_1 z_2; z_3 z_4} = \sum_{x_1 x_2 x_3 x_4}
\overline{c}_{x_1 x_2; x_3 x_4} T_{x_1 z_1}T_{x_2 z_2}T_{x_3 z_3}T_{x_4 z_4}+
b_{z_1 z_2; z_3 z_4},
\label{RG}
\ee
where the matrix $T_{x z}$ is defined in Eq. (28) of  
Ref. \cite{HasenPer} and called there $M(n,n_B)$. It satisfies the 
properties
\begin{eqnarray}
\sum_{x}T_{x z} & = & 4,  \label{obs1}\\
\sum_{z}T_{x z} & = & 1,              \\
T_{xz} & = & T_{x+2n,z+n},            \\
\sum_{x_2 x_3 x_4} b_{x_1 x_2; x_3 x_4} & = & 0,
\label{obs4}
\end{eqnarray}
where $n$ is an arbitrary lattice vector. It is also important to notice that 
$T_{x z}$ is strongly peaked around $(x-2z)_{\mu}\in \{0,1\}$ for
the range of $\kappa$ considered in numerical computations.

The first step in trying to solve Eq. (\ref{RG}) (more precisely the
equation $\overline{c}'=\overline{c}$) is to decide whether to work
in real space or in Fourier space. Since Eq. (\ref{RG}), once written in
Fourier space, does not map continuous functions into continuous functions,
the more convenient choice is the first one.
Moreover by working in real space one can take advantage of locality.

In order to solve Eq. (\ref{RG}) one must face various technical problems.
First of all the equation involves an infinite number of couplings.
However, as noticed by Hasenfratz and Niedermayer \cite{HasenPer}, 
the relevant couplings are of short-range type unless $\kappa$ is very
small or very large. One can thus hope to obtain reasonable 
approximations if one considers truncations that involve couplings
among the spins of nearby points.
We have therefore considered only couplings that are defined in an
$l\times l$ square (a more precise definition will be given below). Notice
that, even for small $l$, the number of couplings turns out to be quite
large. Indeed it increases roughly as $(2l+1)^6$. In particular the 
number of inequivalent couplings is $51$, $771$, and $5329$ respectively
for $l=1,2,3$.

To define the domain precisely, let us begin by noticing that the 
couplings $\overline{c}_{x_1 x_2; x_3 x_4}$ should be 
invariant under the group of lattice symmetries (including of course the 
translations)
which acts on the four points $[x]\equiv (x_1, x_2; x_3, x_4)$
and under the group of permutations generated by
\be
(x_1, x_2; x_3, x_4)\rightarrow \, (x_2, x_1; x_3, x_4) \qquad\qquad
(x_1, x_2; x_3, x_4)\rightarrow \, (x_3, x_4; x_1, x_2).
\ee
To each set of points $[x]$ we associate the quantity
$|[x]|\equiv \max_{\mu,i}\{x_{i,\mu}\}$ and 
a new set of lattice points
$[d\ ] \equiv (d_1, d_2; d_3, d_4)$ that are equivalent to $[x]$ 
under the above symmetries and that satisfy the following two properties:
(i) $d_1 = (0,0)$; (ii)
$|[d]|\le |[y]|$ for all 
$[y]$ equivalent to $[d]$ --- and therefore to $[x]$ --- such that
$y_1 = (0,0)$. The truncation to the $l\times l$ square is defined 
by keeping all couplings $\overline{c}_{x_1 x_2; x_3 x_4}$ such that 
$|[d]|\le l$.

Because of this truncation, solving Eq. \reff{RG} becomes a standard 
linear algebra problem. Eq. \reff{RG} can be rewritten as 
\be
\overline{c}_i = \sum_{j\in \Lambda^l} \mathbb{T}_{ij} \overline{c}_j + b_i,
\label{lintrunc}
\ee
where the indices $i,j$ run over the set $\Lambda^l$ of 
inequivalent (with respect to the above symmetries) 
couplings $\overline{c}_i$ that are defined in an $l\times l$ square.

Notice a further difficulty in Eq. \reff{RG}: in principle we should sum
over $x_i$ ranging on all $\mathbb{Z}^2$. 
However thanks to the observation which follows Eq. (\ref{obs4}), 
the contributions to the sum are very small unless
$|x_{i\mu}|\ltapprox (2l+1)$. Practically we
found $|x_{i\mu}|\le (2l+4)$ to be enough at our level of precision.

Now we have to solve the linear system (\ref{lintrunc}). One could try to
find the solution iterating 
Eq. (\ref{lintrunc}) a sufficient number of times (after all, it is an RG
transformation). Unfortunately this approach does not work: because 
of Eq. (\ref{obs1}), $\mathbb{T}$ has an
eigenvalue equal to $256$ so that the iteration of  Eq. (\ref{lintrunc})
diverges badly. This  eigenvalue is associated to the
operator: $\sum_{x_1 x_2 x_3 x_4}
(\ph_{x_1}\cdot\ph_{x_2}) (\ph_{x_3}\cdot\ph_{x_4})$ (a rather dummy one).\\
A better approach is to solve the equation
\be
(1-\mathbb{T})\overline{c} = b,
\label{tobesolved}
\ee
with a standard method (we used both LU decomposition and singular value
decomposition).\\
The problem is now that $\mathbb{T}$ has two unit eigenvalues\footnote{
A necessary condition for Eq. \reff{tobesolved} to have solutions is that 
$v_i\cdot b = 0$, where $v_1$ and $v_2$ are the left eigenvectors of 
$\mathbb{T}$ with eigenvalue 1. We have explicitly verified this condition.}
which are associated to two marginal operators: 
\begin{eqnarray}
O_1 & = & \int dx\, \left[\ph\cdot\Box\ph\right]^2,\\
O_2 & = & \int dx\, (\ph^a \Box\ph^b) (\ph^a \Box\ph^b).
\end{eqnarray}
In other words,
Eq. \reff{RG} does not determine $\overline{c}_{x_1 x_2; x_3 x_4}$ 
uniquely, but there is the freedom to add two additional conditions.
However there are additional constraints on 
$\overline{c}_{x_1 x_2; x_3 x_4}$ that follow from Eq. (\ref{C4})
and the fact that $w_{xy}$ is given.
For instance, using Eq. (\ref{C4}),
and the fact that the two-spin coupling has the standard normalization
$\sum_x w_{xy}(x-y)^2 = -4$, one can see that the four-spin couplings
should satisfy the conditions
\begin{eqnarray}
\sum_{x_2 x_3 x_4}\overline{c}_{x_1 x_2; x_3 x_4}
          \left[(x_3-x_1)^2+(x_4-x_1)^2 \right] & = & -1,
\label{vinc1}\\
\sum_{x_2 x_3 x_4}\overline{c}_{x_1 x_2; x_3 x_4}
          \left[2(x_3-x_1)\cdot(x_4-x_2)\right] & = & -1.
\label{vinc2}
\end{eqnarray}
One can check that Eqs. \reff{vinc1} and \reff{vinc2} are not 
automatically satisfied by Eq. \reff{RG}. 
Therefore they provide the additional equations necessary to have a unique 
solution to the linear system \reff{lintrunc}. The solution is obtained 
using the singular-value decomposition method. 
We have verified that the solution of the equation \reff{RG} and of the 
two constraints \reff{vinc1} and \reff{vinc2} satisfies Eq. \reff{C4}.

With this procedure we have been able to compute the couplings 
$\overline{c}_{x_1 x_2; x_3 x_4}$ for various values of $\kappa$ and 
for $l=1$ and $l=2$. In order to use the results of appendix
\ref{AppB} we have reexpressed the integrals $A$, $B$,
$D$, and $F$ in terms of real-space 
quantities:
\begin{eqnarray}
&& \hskip -10pt
A =\, \frac{1}{48}\sum_x \left\{G_{x0} w_{x0} \sum_{\mu}x^4_{\mu}\right\},\\
&& \hskip -10pt
B =\, \frac{1}{24}\sum_x G_{x0} w_{x0} x_0^2 x_1^2,\\
&& \hskip -10pt
D =\, \frac{1}{6}\sum_{x_2 x_3 x_4}\left\{c_{x_1 x_2 x_3 x_4}
G_{x_3 x_4}
\left[\frac{1}{2}
\sum_{\mu}(x_2-x_1)^4_{\mu}-3(x_2-x_1)^2_0(x_2-x_1)^2_1\right]\right\},
\label{dtilde} \\
&& \hskip -10pt
F =\, \frac{1}{3}\sum_{x_2 x_3 x_4}\left\{c_{x_1 x_2 x_3 x_4}
\left( G_{x_1 x_3}
-G_{x_1 x_4}-G_{x_2 x_3}+G_{x_2 x_4}\right)\times\right.\nonumber\\
&&\qquad \times\left.\left[\frac{1}{2}
\sum_{\mu}(x_4-x_1)^4_{\mu}-3(x_4-x_1)^2_0(x_4-x_1)^2_1\right]\right\},
\label{ftilde}
\end{eqnarray}
where $G_{xy}$ is the lattice propagator, defined by 
$\sum_z G_{xz} w_{zy} = \delta_{xy}$. 
The final results are reported in Table \ref{qres}.

\begin{table}
\begin{tabular}{|c|c|c|c|c|c|}
\hline
$\kappa$&$(A-3B)$&\multicolumn{2}{|c|}{$D$}&
\multicolumn{2}{|c|}{$F$}\\
\hline
& & $ l = 1 $ & $ l = 2 $ & $ l = 1 $ & $ l = 2 $\\
\hline
\hline
$0.25$  & $-0.041\cdot 10^{-3}$ & $139.1\cdot 10^{-3}$ &
    $204.1\cdot 10^{-3}$ & $-28.79\cdot 10^{-3}$ & $-18.88\cdot 10^{-3}$\\
$0.5$   & $-0.178\cdot 10^{-3}$ & $57.82\cdot 10^{-3}$ &
    $72.95\cdot 10^{-3}$ & $-10.55\cdot 10^{-3}$ &$-6.711\cdot 10^{-3}$\\
$0.625$ & $-0.266\cdot 10^{-3}$ & $32.98\cdot 10^{-3}$ &
    $44.05\cdot 10^{-3}$ & $-3.336\cdot 10^{-3}$ & $-3.912\cdot 10^{-3}$\\
$0.75$  & $-0.364\cdot 10^{-3}$ & $25.17\cdot 10^{-3}$ &
    $26.18\cdot 10^{-3}$ & $-2.555\cdot 10^{-3}$ & $-2.215\cdot 10^{-3}$\\
$1.0$   & $-0.579\cdot 10^{-3}$ & $16.23\cdot 10^{-3}$ &
    $8.292\cdot 10^{-3}$ & $-2.299\cdot 10^{-3}$ & $-0.792\cdot 10^{-3}$\\
$1.5$   & $-1.051\cdot 10^{-3}$ & $5.579\cdot 10^{-3}$ &
    $-0.008\cdot 10^{-3}$& $-1.745\cdot 10^{-3}$ & $-1.131\cdot 10^{-3}$\\
$2.0$   & $-1.541\cdot 10^{-3}$ & $1.563\cdot 10^{-3}$ &
    $-0.510\cdot 10^{-3}$ & $-2.387\cdot 10^{-3}$ & $-1.957\cdot 10^{-3}$\\
$3.0$   & $-2.494\cdot 10^{-3}$ & $6.609\cdot 10^{-3}$ &
    $-3.031\cdot 10^{-3}$ & $-8.247\cdot 10^{-3}$ & $-1.972\cdot 10^{-3}$\\
$4.0$   & $-3.363\cdot 10^{-3}$ & $24.66\cdot 10^{-3}$ &
    $-7.111\cdot 10^{-3}$ & $-20.99\cdot 10^{-3}$ & $-1.292\cdot 10^{-3}$\\
\hline
\end{tabular}
\caption{Results for the integrals appearing in the calculation of the 
mass gap for the perfect action. Here $\kappa$ is the parameter appearing 
in the RG transformations and $l$ refers to the domain in which the couplings
are considered.}
\label{qres}
\end{table}
%

%
%

%
\clearpage

\begin{figure}
\vspace*{-1cm} \hspace*{-0cm}
\begin{center}
\epsfxsize = 0.9\textwidth
\leavevmode\epsffile{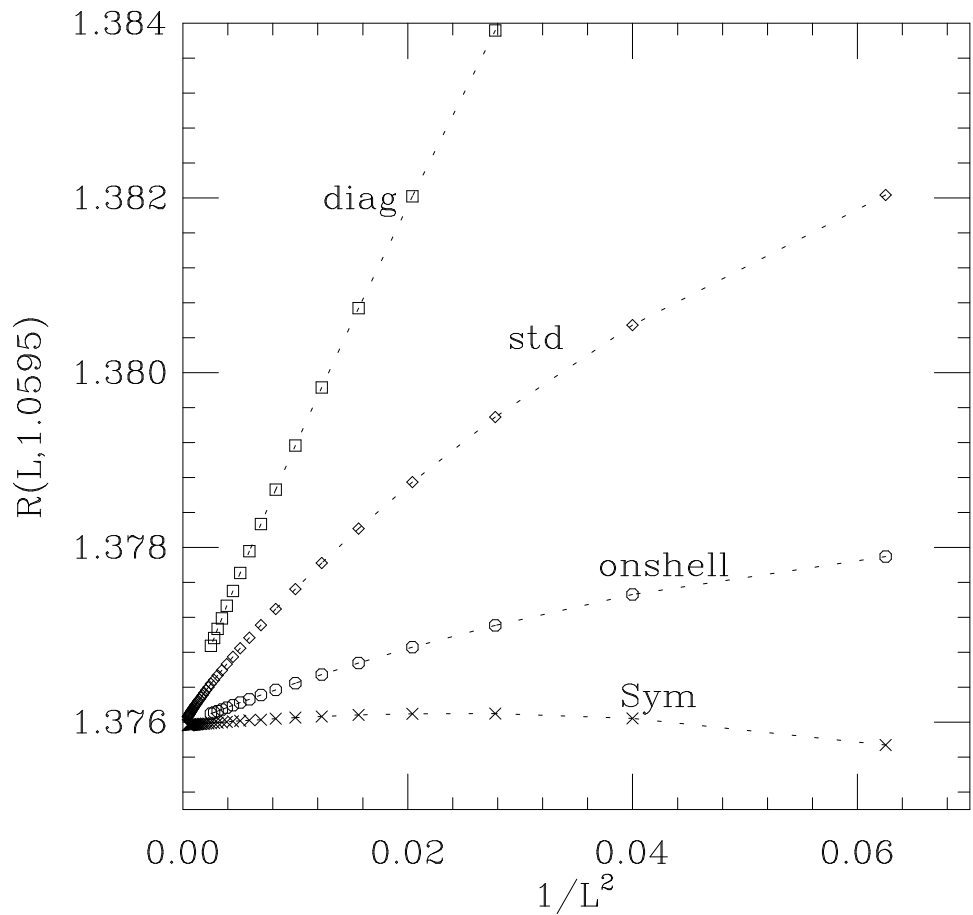}
\end{center}
\vspace*{-1cm}
\caption{$\mu(2L,\b) 2L$ at $N=\infty$ with $\mu(L,\b) L=1.0595$ fixed for the 
various actions introduced in the text. ``Sym", ``onshell", ``std", 
``diag" refer respectively to the actions \protect\reff{sym},
\protect\reff{ons}, \protect\reff{std}, \protect\reff{dia}.}
\label{RNinf}
\end{figure}

\begin{figure}
\vspace*{-1cm} \hspace*{-0cm}
\begin{center}
\epsfxsize = 0.9\textwidth
\leavevmode\epsffile{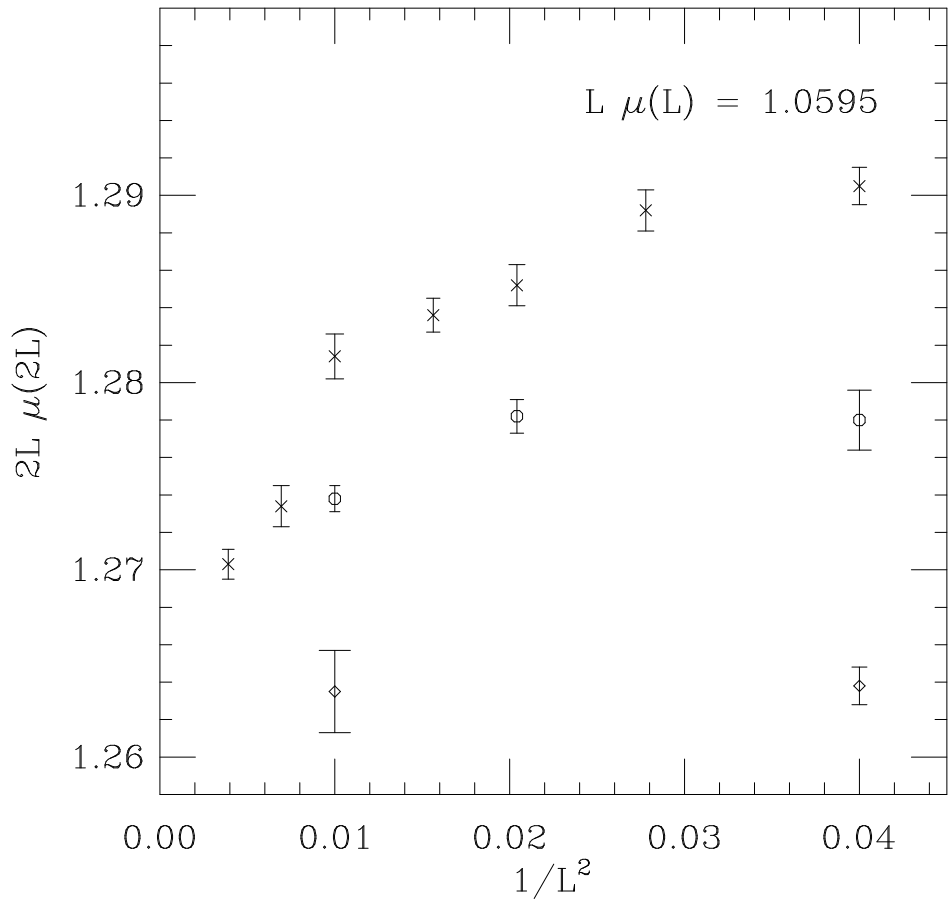}
\end{center}
\vspace*{-1cm}
\caption{$\mu(2L,\b) 2L$ at $N=3$ with $\mu(L,\b) L=1.0595$ fixed.
Crosses refer to the standard action, circles to the on-shell 
tree-level improved action, and diamonds to the perfect action.}
\label{fig3}
\end{figure}

\begin{figure}
\vspace*{-1cm} \hspace*{-0cm}
\begin{center}
\epsfxsize = 0.9\textwidth
\leavevmode\epsffile{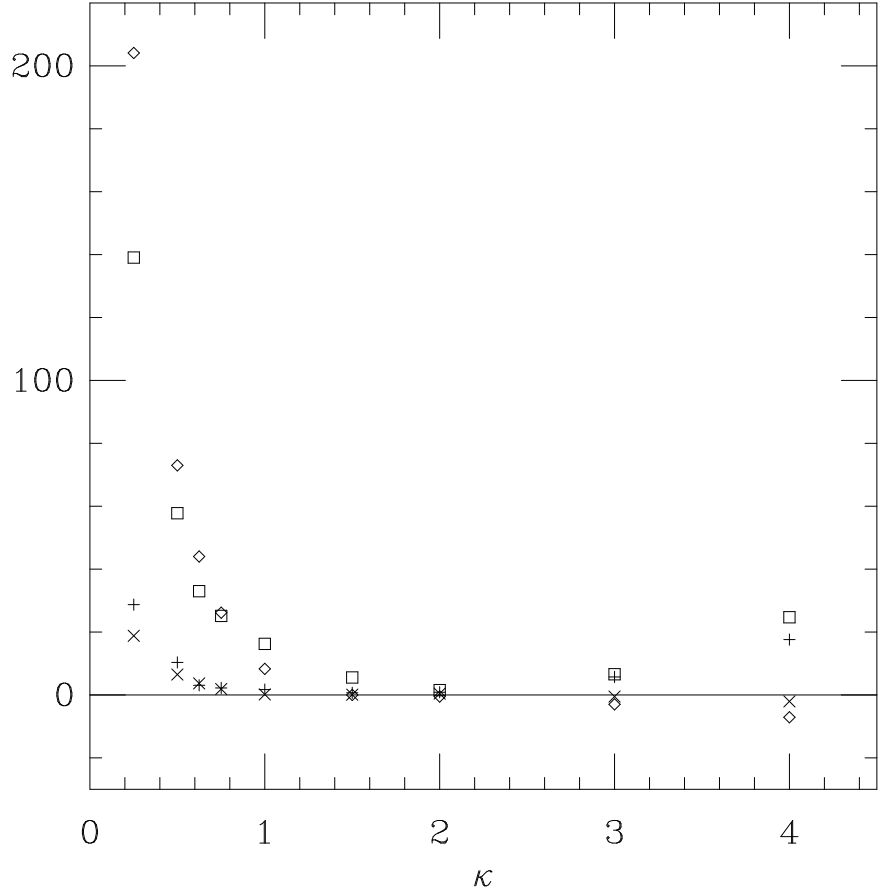}
\end{center}
\vspace*{-1cm}
\caption{One-loop perturbative coefficients for the mass gap 
for spatial momentum ${p}\not=0$. Results for various values of the 
renormalization-group parameter $\kappa$ and truncation index $l$.
Squares and diamonds correspond to $D$ for $l=1$ and $l=2$ respectively.
Pluses and crosses to the combination $A-3B-F$ for $l=1$ and $l=2$ 
respectively. The vertical scale has been multiplied by $10^3$.
}
\label{figmassgap}
\end{figure}

\begin{figure}
\vspace*{-1cm} \hspace*{-0cm}
\begin{center}
\epsfxsize = 0.9\textwidth
\leavevmode\epsffile{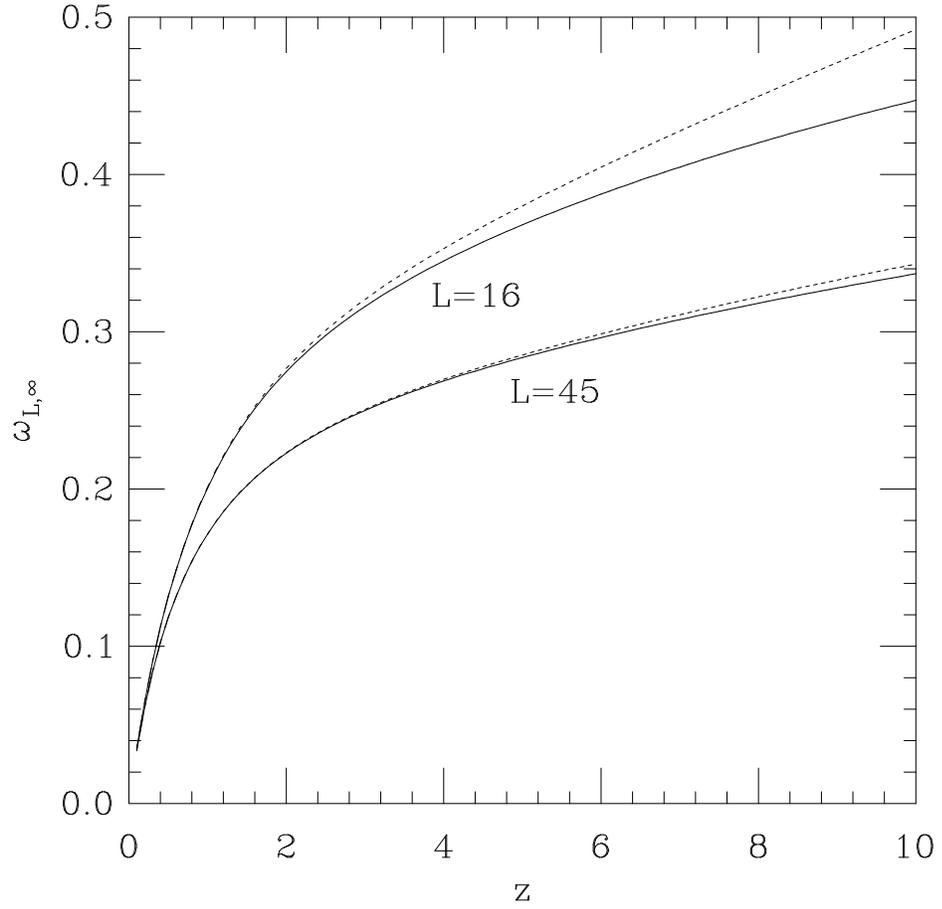}
\end{center}
\vspace*{-1cm}
\caption{$\omega_{L,\infty}$ for two different values of $L$. 
The continuous line is the exact value obtained from the gap equations,
while the dotted line is the solution of the equation
\protect\reff{omegaLT}. }
\label{figomega}
\end{figure}

\begin{figure}
\vspace*{-1cm} \hspace*{-0cm}
\begin{center}
\epsfxsize = 0.9\textwidth
\leavevmode\epsffile{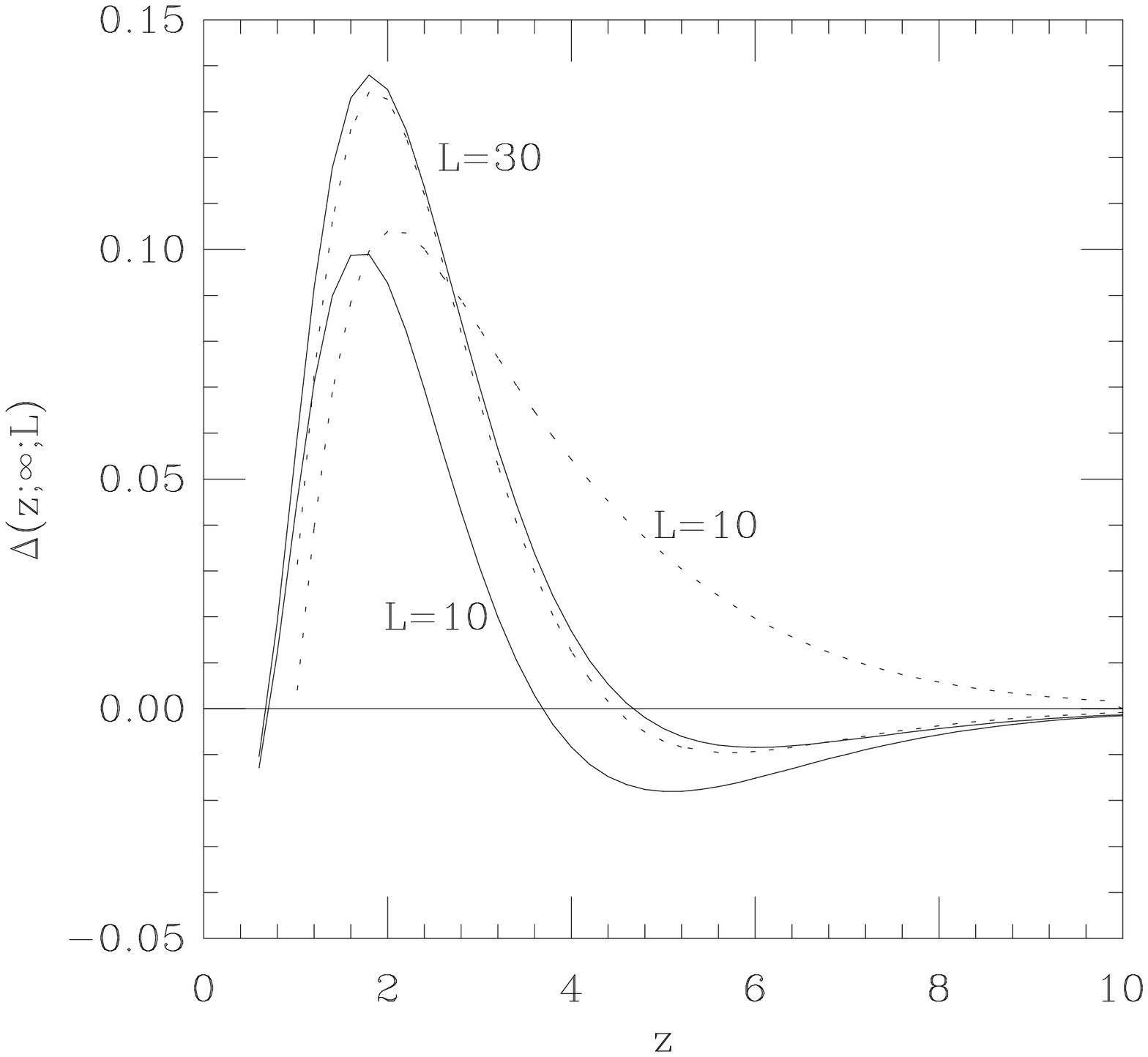}
\end{center}
\vspace*{-1cm}
\caption{Scaling corrections for $\rho=\infty$ and 
two different values of $L$, $L = 10$ and $L=30$.
The full line is the exact expression 
$L^2 (m^2_\infty/m^2_{L,\infty}\times 1/f_m(z,\infty) - 1)$, 
cf. Eq. \protect\reff{espansioneasminfsumLT}, the dashed line is the 
asymptotic expression $\Delta_m(z;\rho;L)$, cf. Eq. \protect\reff{Deltam}.
}
\label{Deltainf}
\end{figure}


\begin{thebibliography}{99}

\bibitem{Symanzik} K. Symanzik, in {\em ``Mathematical problems in
theoretical physics"}, R. Schrader et al. eds., (Springer, Berlin, 1982);
Nucl. Phys. {\bf B226} (1983) 187, 205.

\bibitem{LuschShell} 
M. L\"uscher and P. Weisz,
Commun. Math. Phys. {\bf 97} (1985) 59; Erratum {\bf 98} (1985) 433.

\bibitem{HasenPer} 
P. Hasenfratz and F. Niedermayer, Nucl. Phys. {\bf B414} (1994) 785.

\bibitem{Hasenfratz_94_98}
P. Hasenfratz, 
Nucl. Phys. {B} (Proc. Suppl.) {\bf 34} (1994) 3;
{\bf 63} (1998) 53.

\bibitem{DeGrand-etal}
T. DeGrand, A. Hasenfratz, P. Hasenfratz, and F. Niedermayer,
Nucl. Phys. {\bf B414} (1994) 785.

\bibitem{Bietenholz-etal}
W. Bietenholz and U.-J. Wiese, 
Nucl. Phys. {\bf 464} (1996) 319.

\bibitem{Orginos-etal}
K. Orginos, W. Bietenholz, R. Brower, S. Chandrasekharan, and 
U. J. Wiese, Nucl. Phys. {B} (Proc. Suppl.) {\bf 63} (1998) 904.

\bibitem{Farchioni-etal}
F. Farchioni, P. Hasenfratz, F. Niedermayer, and A. Papa,
Nucl. Phys. {\bf B454} (1995) 638.

\bibitem{Spiegel} S. P. Spiegel, Phys. Lett. {\bf B400} (1997) 352.

\bibitem{Hasenfratz-Niedermayer_97}
P. Hasenfratz and F. Niedermayer,
Nucl. Phys. {\bf B507} (1997) 399.

\bibitem{Luscher_LAT97}
S. Capitani, M. Guagnelli, M. L\"uscher, S. Sint, R. Sommer,
P. Weisz, and H. Wittig,
Nucl. Phys. B (Proc. Suppl.) {\bf 63} (1998) 153 and references therein.

\bibitem{Edwards_LAT97}
R. G. Edwards, U. M. Heller, and T. R. Klassen,
Nucl. Phys. B (Proc. Suppl.) {\bf 63} (1998) 847.

\bibitem{Gockeler_LAT97}
M. G\"ockeler, R. Horsley, H. Oelrich, H. Perlt, P. Rakow,
G. Schierholz, and A. Schiller,
Nucl. Phys. B (Proc. Suppl.) {\bf 63} (1998) 868.

\bibitem{Dawson_LAT97}
C. Dawson, G. Martinelli, G. C. Rossi, C. T. Sachrajda,
S. Sharpe, M. Talevi, and M. Testa,
Nucl. Phys. B (Proc. Suppl.) {\bf 63} (1998) 877.

\bibitem{Polyakov_75}  A. M. Polyakov, Phys. Lett. {\bf B59} (1975) 79.

\bibitem{Brezin_76}  E. Br\'ezin and J. Zinn-Justin,
  Phys. Rev. B {\bf 14} (1976) 3110.

\bibitem{Bardeen_76}
W. A. Bardeen, B. W. Lee, and R. E. Shrock, Phys. Rev. D {\bf 14} (1976) 985.

\bibitem{Stanley}  H.~E. Stanley, Phys. Rev. {\bf 176} (1968) 718;
{\bf 179} (1969) 570.

\bibitem{DiVecchia} P. Di Vecchia, R. Musto, F. Nicodemi, R.~Pettorino,
and P.~Rossi,
  Nucl. Phys. {\bf B235} [FS11] (1984) 478.

\bibitem{Muller}  V. F. M\"uller, T. Raddatz, and W. R\"uhl,
  Nucl. Phys. {\bf B251} [FS13] (1985) 212;
  {\em erratum} Nucl. Phys. {\bf B259} (1985) 745.

\bibitem{Flyvbjerg}  H. Flyvbjerg, Phys. Lett. {\bf B219} (1989) 323;
  H. Flyvbjerg and S. Varsted, Nucl. Phys. {\bf B344},
  (1990) 646;
  H. Flyvbjerg and F. Larsen, Phys. Lett. {\bf B266} (1991) 92, 99.

\bibitem{Campostrini_90ab}  P. Biscari, M. Campostrini, and P. Rossi,
  Phys. Lett. {\bf B242} (1990) 225;
  M. Campostrini and P. Rossi, Phys. Lett. {\bf B242} (1990) 81.

\bibitem{Zamolodchikov_79}  A. B. Zamolodchikov and A. B. Zamolodchikov,
  Nucl. Phys. {\bf B133} (1978) 525;
  Ann. Phys. (N.Y.) {\bf 120} (1979) 253.

\bibitem{Polyakov-Wiegmann_83}  A. Polyakov and P. B. Wiegmann,
  Phys. Lett. {\bf 131B} (1983) 121.

\bibitem{Hasenfratz-Niedermayer_1}  P. Hasenfratz, M. Maggiore, and
  F. Niedermayer, Phys. Lett. {\bf B245} (1990) 522.

\bibitem{Hasenfratz-Niedermayer_2} P. Hasenfratz and F. Niedermayer,
  Phys. Lett. {\bf B245} (1990) 529; {\bf B268} (1991) 231.

\bibitem{Wolff_O4_O8}  U. Wolff, Phys. Lett. {\bf B248} (1990) 335.

\bibitem{MGMC_O4}  R. G. Edwards, S. J. Ferreira, J. Goodman,
and A. D. Sokal,
  Nucl. Phys. {\bf B380} (1992) 621.

\bibitem{o3_letter} S. Caracciolo, R. G. Edwards,
   A. Pelissetto, and A. D. Sokal, Phys. Rev. Lett. {\bf 75} (1995) 1891;
  Nucl. Phys. B (Proc. Suppl.) {\bf 42} (1995) 752.

\bibitem{CEMPS}
S.~Caracciolo, R.~G.~Edwards, T.~Mendes, A.~Pelissetto, and A.~D.~Sokal,
Nucl. Phys. B (Proc. Suppl.) {\bf 47} (1996) 763.

\bibitem{Alles-Symanzik}
B. All\'es, A. Buonanno, and G. Cella,
Nucl. Phys. {\bf B500} (1997) 513;
Nucl. Phys. B (Proc. Suppl.) {\bf 53} (1997) 677.

\bibitem{Alles-e-turchi}
B. All\'es, G. Cella, M. Dilaver, and Y. Gunduc,
{\em Testing fixed points in the $2D$ $O(3)$ non-linear 
$\sigma$-model}, {\tt hep-lat/9808003}.

\bibitem{Falcioni-Treves}
M. Falcioni and A. Treves, Nucl. Phys. {\bf B265} (1986) 671.

\bibitem{CP-3loop} S. Caracciolo and A. Pelissetto,
Nucl. Phys. {\bf B420} (1994) 141.

\bibitem{CP-4loop} S. Caracciolo and A. Pelissetto,
Nucl. Phys. {\bf B455} [FS] (1995) 619.

\bibitem{Shin} 
D.-S. Shin, {\em Correction to four-loop RG functions in the two-dimensional
lattice $O(n)$ $\sigma$-model}, {\tt hep-lat/9810025}.

\bibitem{Caracciolo-Pelissetto_97} 
S. Caracciolo and A. Pelissetto, Phys. Lett. {\bf B402} (1997) 305.

\bibitem{Wolff_89_90} U. Wolff,
Phys. Rev. Lett. {\bf 62} (1989) 361;
Nucl. Phys. {\bf B322} (1989) 759; {\bf B344} (1990) 581.

\bibitem{Edwards-Sokal_89}
R. G. Edwards and A. D. Sokal, Phys. Rev. D {\bf 40} (1989) 1374.

\bibitem{Hasenbusch_90}
M. Hasenbusch, Nucl. Phys. {\bf B333} (1990) 581.

\bibitem{Caracciolo-etal_93}
S. Caracciolo, R. G. Edwards, A. Pelissetto, and A. D. Sokal,
Nucl. Phys. {\bf B403} (1993) 475;
Nucl. Phys. B (Proc. Suppl.) {\bf 20} (1991) 72; {\bf 26} (1992) 595.

\bibitem{Buonanno-Cella_95}
A. Buonanno and G. Cella, Phys. Rev. D {\bf 51} (1995) 5865.

\bibitem{Caracciolo-etal_98} 
S. Caracciolo, A. Montanari, and A. Pelissetto,
Nucl. Phys. B (Proc. Suppl.) {\bf 63} (1998) 916.

\bibitem{LuschNum} M. L\"uscher, P. Weisz, and U. Wolff,
Nucl. Phys. {\bf B359} (1991) 221.

\bibitem{Shin_97}
D.-S. Shin, Nucl. Phys. {\bf B496} (1997) 408.

\bibitem{Niedermayer_LAT96} F. Niedermayer,
Nucl. Phys. B (Proc. Suppl.) {\bf 53} (1997) 56.

\bibitem{Bell-Wilson_74}
T. L. Bell and K. Wilson,
Phys. Rev. B {\bf 10} (1974) 3935.

\bibitem{Caracciolo-Pelissetto_98} 
S. Caracciolo and A. Pelissetto, 
Phys. Rev. D {\bf 58} (1998) 105007.

\bibitem{Magnoli-Ravanini}  N. Magnoli and F. Ravanini,
  Z. Phys. {\bf C34} (1987) 43.   

\bibitem{Luscher_81}
M. L\"uscher, Phys. Lett. {\bf 118B} (1982) 391.

\bibitem{Swendsen-Wang}
R. H. Swendsen and J.-S. Wang, 
Phys. Rev. Lett. {\bf 58} (1987) 86.

\bibitem{Edwards-Sokal_88}
R. G. Edwards and A. D. Sokal, Phys. Rev. D {\bf 38} (1988) 2009.

\bibitem{Alles_unpublished}
B. All\'es, S. Caracciolo, G. Cella, M. D'Elia, A. Montanari, and 
A. Pelissetto, unpublished.

\bibitem{BF-Arch-Rat-Mech-Anal}
M. E. Fisher and M. N. Barber,
Arch. Rat. Mech. Anal. {\bf 47} (1972) 205.

\bibitem{RV} P. Rossi and E. Vicari,
Phys. Lett. {\bf B389} (1996) 571.

\bibitem{Montvay} 
B. Berg, S. Meyer, I. Montvay, and K. Symanzik,
Phys. Lett. {\bf 126B} (1983) 467.


\end{thebibliography}
\end{document}